\documentclass[a4paper,fleqn,usenatbib]{mnras}
\usepackage[T1]{fontenc}
\usepackage{ae,aecompl}
\usepackage{graphicx}
\usepackage{amssymb}
\usepackage{amsmath}
\usepackage{subfigure,float,verbatim,placeins,array, tabularx,tabulary,tabu,booktabs,xtab,makecell}
\usepackage{ifpdf}
\usepackage{threeparttable}
\usepackage{gensymb}
\usepackage{mathrsfs}

\def\aap{A\&A}
\def\apjl{ApJL}
\def\apjs{ApJS}

\title[Validating stellar dynamical models with CO]{The EDGE-CALIFA survey: Validating stellar dynamical mass models with CO kinematics}
\author[Leung et al.]{Gigi Y.\,C. Leung$^{1}$\thanks{Email: leung@mpia.de}, Ryan Leaman$^{1}$, Glenn van de Ven$^{1,2}$, Mariya Lyubenova$^{2}$,
\and Ling Zhu$^{1}$, Alberto D. Bolatto$^{3}$, Jesus Falc\'{o}n-Barroso$^{4,5}$, Leo Blitz$^{6}$, 
\and Helmut Dannerbauer$^{4,5}$, David B. Fisher$^{7}$, Rebecca C. Levy$^{3}$, Sebastian F. Sanchez$^{8}$, 
\and Dyas Utomo$^{6,9}$, Stuart Vogel$^{3}$, Tony Wong$^{10}$ and Bodo Ziegler$^{11}$\\
$^{1}$Max-Planck Institut f\"ur Astronomie, K\"onigstuhl 17, D-60117 Heidelberg, Germany\\
$^{2}$European Southern Observatory, Karl-Schwarzschild-Str. 2, D-85741 Garching, Germany\\
$^{3}$Department of Astronomy, University of Maryland, College Park, MD 20742, USA\\
$^{4}$Instituto de Astrof\'isica de Canarias, C/ Via L\'actea s/n, 38200 La Laguna, Canary Islands. Spain\\ 
$^{5}$Departamento de Astrof\'isica, Universidad de La Laguna (ULL), E-38206 La Laguna, Tenerife, Spain \\ 
$^{6}$Department of Astronomy, University of California, Berkeley, CA 94720, USA\\
$^{7}$Centre for Astrophysics and Supercomputing, Swinburne University of Technology, Hawthorn, Victoria 3122, Australia\\
$^{8}$Instituto de Astronomía, Universidad Nacional Autonóma de Mexico, A.P. 70-264, 04510 México, D.F., Mexico\\
$^{9}$Department of Astronomy, The Ohio State University, Columbus, OH 43210, USA\\
$^{10}$Department of Astronomy, University of Illinois, Urbana, IL 61801, USA\\
$^{11}$University Vienna, Turkenschanzstraße 17,1180 Wien, Austria}
\date{Accepted 2018. Received 2017; in original form 2017 }

\pagerange{\pageref{firstpage}--\pageref{lastpage}} \pubyear{2018}

\begin{document}

\label{firstpage}

\maketitle

\begin{abstract}

Deriving circular velocities of galaxies from stellar kinematics can provide an estimate of their total dynamical mass, provided a contribution from the velocity dispersion of the stars is taken into account.  Molecular gas (e.g., CO) on the other hand, is a dynamically cold tracer and hence acts as an independent circular velocity estimate without needing such a correction.  In this paper we test the underlying assumptions of three commonly used dynamical models, deriving circular velocities from stellar kinematics of 54 galaxies (S0-Sd) that have observations of both stellar kinematics from the CALIFA survey, and CO kinematics from the EDGE survey.  We test the Asymmetric Drift Correction (ADC) method, as well as Jeans, and Schwarzschild models.  The three methods each reproduce the CO circular velocity at 1$R_\mathrm{e}$ to within 10\%.  All three methods show larger scatter (up to 20\%) in the inner regions ($R<0.4R_\mathrm{e}$) which may be due to an increasingly spherical mass distribution (which is not captured by the thin disk assumption in ADC), or non-constant stellar M/L ratios (for both the JAM and Schwarzschild models). This homogeneous analysis of stellar and gaseous kinematics validates that all three models can recover $M_\mathrm{dyn}$ at 1$R_\mathrm{e}$ to better than 20\%, but users should be mindful of scatter in the inner regions where some assumptions may break down.
\end{abstract}

\begin{keywords}

galaxies: kinematics and dynamics - galaxies: spiral.

\end{keywords}

\section{Introduction}\label{sect_intro}

The kinematics of stars or gas in galaxies allows one to trace its underlying gravitational potential and hence the enclosed mass within a particular radius. In particular, the circular velocity, $V_\mathrm{c}$, defined as $V_\mathrm{c}^{2}(R)\equiv-R(\partial\Phi/\partial R)$, is an optimal tracer of a galaxy's potential. The mass profile of galaxies provides insight into, for example, understanding how baryons and dark matter co-habitate in galaxies, how the galaxies assemble, and how galaxy evolution proceeds across the Hubble sequence in a variety of environments \citep[e.g. see reviews:][and references therein]{cou14,cap16}. 

Typical kinematic tracers for galaxies include atomic, molecular or ionised gas, and stars. While observations of stellar kinematics can be done at high spatial resolution, the high velocity dispersion intrinsic to the stellar component renders their dynamical analysis non-trivial. Luminous ionised gas can be similarly complicated due to turbulent shocks surrounding star formation (of which it is associated). Molecular gas, such as the CO, which is often used as a tracer of H$_\mathrm{2}$, typically is dynamically cold with an intrinsic dispersion of $\sim10\mathrm{\,km\,s^{-1}}$ at low redshift \citep{mog16}, meaning the rotation curves closely follow the circular velocities and therefore is an optimal tracer of $V_\mathrm{c}$.  However, molecular gas is found in the disk plane and can often show kinematic features due to perturbations occurring in the disk by a bar or spiral arms \citep[e.g.][]{lai99,she07}. A method for removing these perturbation, for example by fitting tilted rings or by harmonic decomposition \citep[e.g.][]{beg87,wong04}, is therefore necessary in order to extract a smooth rotation curve from molecular gas. 

As stars are collisionless, their orbits can cross and stars born from the cold molecular gas eventually dynamically evolve to have large random motions at present day (Leaman et al. 2017), resulting in a velocity dispersion up to a hundred $\mathrm{km\,s^{-1}}$ at the effective radius and  as high as a few hundreds $\mathrm{km\,s^{-1}}$ in the galactic bulge. Hence when deriving a circular velocity from stellar kinematics, we must take into account both the rotation velocity and the velocity dispersion, especially when they are of comparable magnitude. There are various methods to recover $V_\mathrm{c}$ from stellar kinematics, typically either by solving the Jeans equations \citep[e.g][]{jeans22,bin90,em94}, by using orbit-based models such as the Schwarzschild model \citep[e.g.][]{sch79,vdM98,tho04,val04}, or by particle-based models such as the Made-to-measure method \citep[e.g.][]{deL07,long10,sye96,zhu14}. Having only the line-of-sight information of the velocity field implies that some assumptions must be made. The Jeans models often make assumptions on, for example, the geometry of the underlying potential, the mass-to-light ratio and the velocity anisotropy profile of the galaxies. Schwarzschild or made-to-measure models, on the other hand, do not make assumption on the velocity anisotropy, but may still require assumptions on the geometry of the gravitational potential and the mass-to-light ratio. 

This paper aims to verify commonly used models in solving for circular velocities from stellar kinematics and calibrate how well each model works in different regimes (e.g. over different radii or galactic properties). We do so by comparing the circular velocities derived from stellar kinematics to those extracted from the molecular gas CO. As the molecular gas and the stars from the same galaxy orbit in the same gravitational potential, the $V_\mathrm{c}$ inferred from their kinematics should match each other. We include three commonly used stellar dynamical models in this study: (1) the asymmetric drift correction (ADC) \citep[e.g. \S4.8][]{bin87,wei08}, (2) the Axisymmetric Jeans Anisotropic Multi-gaussian expansion (JAM) model \citep{cap08}, and (3) the orbit-based Schwarzschild model \citep{sch79,van08}. Both ADC and JAM derive $V_\mathrm{c}$ by solving the Jeans equations. Among the three, ADC is the most simplistic model and assumes that stars lie on a thin disk with either a constant or a parametrised form of velocity anisotropy. JAM removes the thin-disk assumption and takes into account the full line-of-sight integration of the stellar kinematics, but still makes assumptions about the velocity anisotropy and the shape of the velocity ellipsoid. The triaxial Schwarzschild models we utilise in this paper are the state of the art in stellar dynamical modelling. The Schwarzschild method is an orbit-based model which does not require any assumption on the shape of velocity ellipsoid, but is expensive in terms of computational power. By comparing the $V_\mathrm{c}$ derived from these three models with that from CO kinematics, we aim to show if and how the relaxation in assumptions allowed by improved computational power in stellar dynamical modelling leads to better constraints in circular velocities. In the remaining of the paper we shall refer the circular velocities derived from CO, ADC, JAM and Schwarzschild models as $V_\mathrm{CO}$, $V_\mathrm{ADC}$, $V_\mathrm{JAM}$ and $V_\mathrm{SCH}$ respectively.

Gas and stellar kinematics have been compared in some individual cases, or for particular applications \citep[e.g.][]{wei08,lea12,bas14,piz04,joh12,hun02}. In particular, \cite{dav13} show, for a sample of 16 early type galaxies (ETG) from the ATLAS$^{3D}$ survey, the agreement of CO and stellar kinematics. Over the late type galaxies, however, a large scale homogeneous test of stellar dynamical models with \emph{cold} gas circular velocity curves is still needed.  

The EDGE \citep{bol17} and CALIFA surveys \citep{san16} respectively provides CO and stellar kinematics over an overlapping sample of nearby galaxies, allowing us to compare the CO rotation curves and stellar circular velocities over a large and homogeneous sample for the first time. Moreover, our sample includes 54 galaxies from type S0 to Scd, allowing us to look for systematic differences in the kinematic tracers as a function of galaxy morphological type. The CALIFA survey also provides H$\alpha$ kinematics; for a comparison between $V_\mathrm{CO}$ and the rotation curves extracted from H$\alpha$ kinematics please refer to Levy et al. (in prep.). 

Readers interested in the data sample may refer to section \ref{sect_data}. In section \ref{sect_method}, we describe the extraction of rotation curves from the CO velocity field. In section \ref{sect_star}, we describe the methods and the underlying assumptions of the three stellar dynamical models (ADC, JAM and Schwarzschild) and compare the circular velocities extracted from stars using different models in section \ref{sect_star}. In section \ref{sect_costar}, we compare the circular velocities extracted using gaseous and stellar kinematics and we characterise the comparisons as functions of radii, local stellar $V/\mathrm{\sigma}_\mathrm\star$ values and galactic parameters. In section \ref{sect_diss}, we discuss the plausible causes for the differences we see. We summarise in section \ref{sect_sum}.

\section{Data}\label{sect_data}
The CARMA Extragalactic Database for Galaxy Evolution (EDGE) survey consist of interferometric observations of 126 galaxies, all of which are included in the Calar Alto Legacy Integral Field Area (CALIFA) survey. The data were obtained using the Combined Array for Research in Millimeter-wave Astronomy (CARMA) at Owens Valley Radio Observatory. These 126 galaxies were mapped in $^{12}$CO($J=1-0$) using the D and E array configuration. Each galaxy typically had 4.3 hours of observation, and all galaxies were observed in the period from December 2014 to April 2015. The velocity resolution of the observations was $20\mathrm{\,km\,s^{-1}}$, and the typical beam has FWHM$_\mathrm{beam}$ of $\sim3-5"$. In this paper, we utilise the integrated intensities, mean velocities and velocity dispersion maps. The kinematic maps are obtained by fitting a gaussian to the spectrum observed at each pixel, with the peak of the fitted gaussian and the standard deviation representing the mean velocity and the velocity dispersion respectively. Complete details of the observations and reduction for the survey, as well as all the CO moment maps, can be found in \citet{bol17}\footnote{the publicly available data can be downloaded from http://www.astro.umd.edu/EDGE/}. 

We obtain stellar kinematics from the CALIFA survey. The observations have a spatial resolution with a FWHM of $\sim2.7"$. The stellar kinematics come from the V1200 data set \citep{fal17}\footnote{the stellar kinematics can be found in http://califa.caha.es}, with a velocity resolution of $\sigma\sim70$\,km\,s$^{-1}$. While the optical and radio community often follow different velocity conventions when extracting kinematics from the observed spectra, both the CO and stellar kinematical maps presented in this paper are converted to: $V\equiv c\Delta\ln\lambda$, where $V$ is the extracted velocities, $c$ is the speed of light and $\Delta\lambda$ is the difference between the observed wavelength and the rest wavelength of any particular lines. This is done to avoid any systematic differences due to the different conventions when comparing the circular velocities extracted from CO and stellar kinematics.

Out of the 126 overlapping galaxies, we select 54 galaxies that provide sufficient signal to noise in CO for us to trace the galaxy kinematics. We select only the galaxies from which we can extract at least three rotation velocity measurements (more on selection criterion in Section \ref{subsect_uncer}). We also exclude merging galaxies as identified for the CALIFA sample in \citet{bar15}, of which the interaction may complicate the differences between gaseous and stellar kinematics. The 54 galaxies of our sample and their parameters as adopted in the CALIFA survey, including the total stellar mass ($M_\star$), distance, inclination ($i$) and photometric position angle ($PA_\mathrm{morph}$) are listed in Table \ref{tab_galpar}.


\begin{table*}
\label{tab_galpar}
\begin{threeparttable}
\begin{tabular}{@{}llllllllll}
\hline
Galaxy & Type\tnote{a} & Distance (Mpc)\tnote{a} & Inclination ($^{\circ}$)\tnote{a} &  log $M_\mathrm\star/M_\mathrm{\odot}$\tnote{a} & $PA_\mathrm{morph}$ ($^{\circ}$)\tnote{a} & $PA_\mathrm{kin}$ ($^{\circ}$)\tnote{b} & $R_\mathrm{e}$ (\arcsec) \tnote{a} & $\sigma_\mathrm\star(R=R_\mathrm{e})$\tnote{a} & $V_\mathrm{sys}$ ($\mathrm{km\,s^{-1}}$)\tnote{b}\\
\hline
\hline

IC0480&		Sc&		65.2&	76.7&	10.15&	167.9&$-$12.1&	24.16&	90.27&	4541.97\\
IC0944&		Sab&	100.0&	69.5&	11.26&	105.7&$-$74.3& 	19.01&	195.23&	6845.50\\
IC1199&		Sb&		60.7&	64.5&	10.60&	157.3&157.3&		20.99&	146.53&	4666.25\\
IC1683\tnote{*}&		Sb&		69.3&	50.5&	10.59&	15.6&$-$161.3&	13.07&	114.72&	4773.18\\
IC2247&		Sab&	61.8&	77.8&	10.51&	148.5&148.5&		21.38&	111.76&	4215.30\\
IC2487&		Sc&		62.2&	78.0&	10.40&	162.9&162.9&		25.34&	116.64&	4278.08\\
IC4566&		Sb&		79.7&	53.9&	10.95&	161.0&$-$19.0& 	15.84&	144.33&	5515.33\\
NGC0477&	Sbc&	83.8&	60.5&	10.48&	128.4&$-$51.6&	21.78&	104.86&	5731.99\\
NGC0496&	Scd&	85.9&	57.0&	10.41&	32.7&$-$147.3&	19.01&	77.23&	5908.02\\
NGC0551&	Sbc&	74.3&	64.2&	10.64&	137.5&137.5&		19.80&	115.83&	5094.57\\
NGC2253&	Sbc&	51.2&	48.3&	10.52&	109.7&109.7&		15.44&	103.49&	3516.41\\
NGC2347&	Sbc&	63.6&	50.7&	10.94&	9.1&	9.1&			18.61&	175.43&	4359.36\\
NGC2410\tnote{*}&	Sb&		66.8&	71.6&	10.88&	34.6&54.6&		21.38&	175.55&	4600.08\\
NGC2639&	Sa&		46.5&	50.2&	11.17&	130.3&130.3&		17.42&	203.92&	3149.31\\
NGC2906&	Sbc&	30.6&	55.7&	10.39&	82.6&82.6&		19.40&	117.05&	2120.89\\
NGC3815&	Sbc&	53.0&	60.4&	10.35&	67.8&$-$112.2&	14.26&	115.52&	3660.81\\
NGC3994&	Sbc&	44.1&	59.9&	10.42&	8.1&8.1& 			9.50&	164.85&	3085.49\\
NGC4047&	Sbc&	49.2&	42.9&	10.69&	97.6&$-$82.4&		16.63&	109.69&	3405.28\\
NGC4149&	Sa&		44.0&	66.2&	10.36&	85.4&$-$94.6&		18.61&	137.22&	3062.19\\
NGC4210&	Sb&		38.8&	41.8&	10.29&	97.7&97.7&		21.38&	83.34&	2701.64\\
NGC4644&	Sb&		70.5&	72.9&	10.45&	57.0&$-$123.0&	12.60&	115.77&	4870.88\\
NGC4711&	Sbc&	58.2&	58.5&	10.31&	41.4&41.4&		17.82&	86.55&	4019.40\\
NGC4961&	Scd&	36.6&	47.4&	9.68&	100.6&$-$79.4&	15.05&	66.73&	2539.42\\
NGC5016&	Sbc&	37.3&	40.9&	10.24&	57.4&$-$122.6&		17.82&	87.83&	2595.19\\
NGC5056&	Sc&		79.4&	55.7&	10.48&	3.4&3.4&	15.84&	120.70&	5494.81\\
NGC5218&	Sab&	41.4&	31.7&	10.65&	101.4&101.4&		18.61&	152.12&	2878.92\\
NGC5480\tnote{*}&	Scd&	27.0&	36.2&	10.14&	41.9&$-$2.0&		25.74&	62.98&	1887.88\\
NGC5520&	Sbc&	26.8&	59.4&	9.86&	63.1&63.1&		12.28&	99.89&	1877.19\\
NGC5633&	Sbc&	33.2&	42.8&	10.26&	16.9&$-$163.1&	13.86&	90.54&	2322.21\\
NGC5784\tnote{*}&	S0&		77.6&	36.4&	11.22&	19.2&$-$110.0&	13.46&	203.65&	5353.01\\
NGC5908\tnote{*}&	Sa&		47.8&	50.7&	11.22&	154.0&$-$27.1&	 34.45&	198.41&	3298.51\\
NGC5980&	Sbc&	58.4&	66.2&	10.72&	11.7&$-$168.3&	17.42&	150.54&	4044.62\\
NGC6060&	Sb&		64.5&	64.3&	10.93&	102.0&$-$78&	 	28.51&	143.28&	4395.10\\
NGC6168&	Sc&		36.8&	76.7&	9.86&	110.2&110.2&		26.93&	66.35&	2523.05\\
NGC6186&	Sb&		42.1&	40.5&	10.57&	49.6&$-$110.2&	20.20&	90.26&	2946.99\\
NGC6301&	Sbc&	119.7&	52.8&	11.02&	108.5&108.5&		24.55&	147.97&	8134.66\\
NGC6394&	Sbc&	121.7&	68.9&	10.90&	42.6&42.6&		14.65&	143.59&	8324.46\\
NGC6478&	Sc&		97.1&	68.4&	11.01&	34.2&$-$145.8&	23.36&	180.19&	6637.44\\
UGC00809&	Scd&	60.0&	79.0&	9.69&	23.6&$-$156.4&	20.20&	93.37&	4146.13\\
UGC03539&	Sc&		47.2&	72.1&	9.85&	117.9&117.9&		20.99&	NaN&	3263.57\\
UGC03969&	Sb&		117.8&	77.4&	10.68&	134.3&$-$45.7&	15.05&	130.62&	7926.60\\
UGC04029&	Sc&		63.5&	77.7&	10.33&	63.5&$-$116.5&	26.14&	108.85&	4350.80\\
UGC04132&	Sbc&	74.4&	72.0&	10.77&	27.6&27.6&		22.97&	160.56&	5092.49\\
UGC05108&	Sb&		116.1&	66.1&	10.89&	138.1&$-$41.9&	9.50&	194.46&	7935.95\\
UGC05598&	Sb&		80.3&	74.8&	10.23&	35.6&35.6&		15.84&	85.51&	5522.98\\
UGC08107&	Sa&		118.6&	71.4&	11.07&	53.2&53.2&		16.63&	207.94&	8086.28\\
UGC09067&	Sbc&	112.1&	62.5&	10.58&	12.6&$-$167.4&	14.65&	139.51&	7649.28\\
UGC09537&	Sb&		126.1&	78.0&	11.22&	140.7&$-$39.3&	20.20&	210.31&	8545.50\\
UGC09542&	Sc&		78.4&	72.7&	10.32&	34.3&34.3&		21.38&	99.09&	5372.05\\
UGC09665&	Sb&		36.5&	74.1&	10.00&	138.2&$-$41.8&	18.61&	74.93&	2568.17\\
UGC09892&	Sbc&	81.1&	72.2&	10.30&	101.0&$-$79.0&	16.63&	73.22&	5564.49\\
UGC10123&	Sab&	53.8&	77.1&	10.52&	53.6&53.6&		18.22&	122.92&	3729.22\\
UGC10384&	Sb&		70.7&	74.3&	10.27&	92.8&92.8&		11.88&	112.19&	4894.88\\
UGC10710&	Sb&		119.7&	69.6&	10.99&	147.2&147.2&		20.20&	168.86&	8144.29\\
\hline
\end{tabular}
\caption{Galaxy sample and their galactic parameters.}
  \begin{tablenotes}
    \item[a] Refer to the CALIFA survey \citep{san16} for the derivation of these values.
    \item[b] From fitting the CO kinematics in this work. $PA_\mathrm{kin}$ denotes the receding side.
    \item[*] $PA_\mathrm{kin}$ as a free parameter when fitting for $V_\mathrm{CO}$ (i.e. $PA_\mathrm{kin}\ne PA_\mathrm{morph}$).
  \end{tablenotes}
\end{threeparttable}
\end{table*}

\section{Extraction of the CO Rotation Velocities and Dispersion Profiles}\label{sect_method}


\begin{figure*}
\includegraphics[trim=320 50 50 50,width=0.32\textwidth,angle=90]{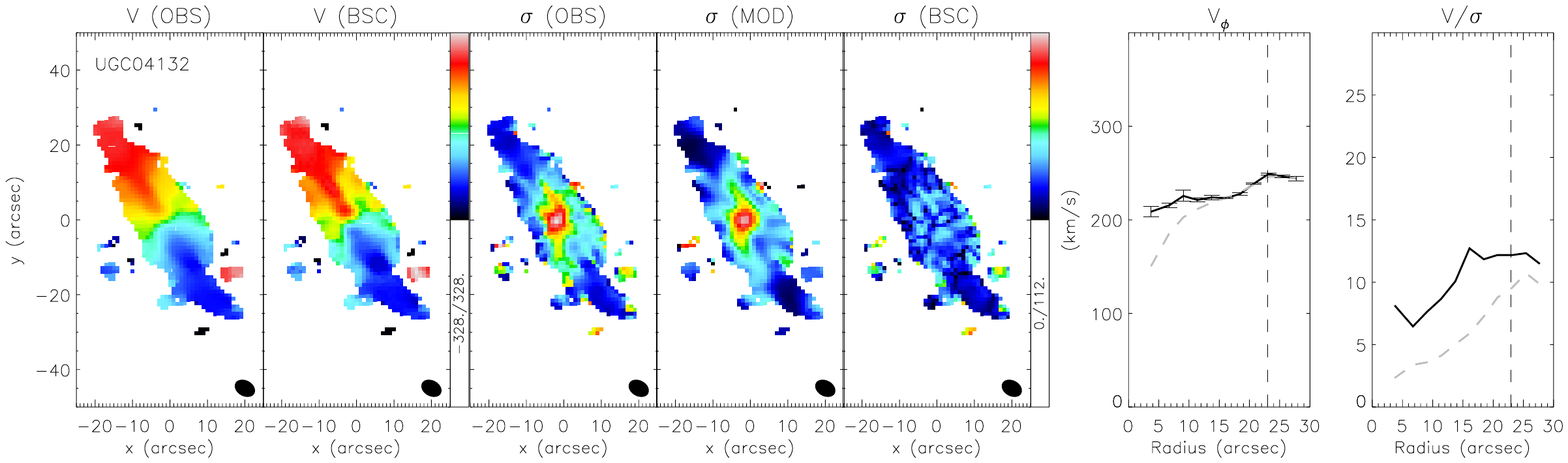}
\caption{Kinematic maps for UGC04132 from the EDGE CO survey. The coloured maps from left to right are: observed mean velocity map, beam-smearing corrected mean velocity map, observed dispersion map, modelled dispersion map and beam-smearing corrected dispersion map. The two plots on the right are the extracted rotation curve and $V/\sigma$ ratio, with the grey dashed line indicating the observed value and black solid line indicate the beam-smearing corrected value which we adopt in our analysis. The black dashed lines mark the effective radius $R_\mathrm{e}$.}
\label{fig_co_map}
\end{figure*}

Before extracting the rotation curves and dispersion profiles, we first applied a beam-smearing correction to both the CO mean velocity and velocity dispersion maps. We describe the beam-smearing correction method in Appendix \ref{app_beamcorr}. In Figure \ref{fig_co_map}, we show as an example the pre- and post- beam-smearing corrected mean velocity and velocity dispersion maps of UGC04132. While we only compare the CO and stellar $V_\mathrm{c}$ beyond 3$\sigma_\mathrm{beam}$ where the effect of beam smearing on the CO rotation curve is insignificant, such a correction to the CO velocity dispersion map is of particular importance in assuring that CO acts as a dynamically cold tracer for our galaxy sample (see Section \ref{subsect_coadc}).

\subsection{Rotation curves}\label{subsect_vrot}
We extract the rotation curves from the CO mean velocity map of each galaxy by first fitting ellipses to the mean velocity map, stepping outwards along the semi-major axes, each time determining the kinematic centre and systemic velocity ($V_\mathrm{sys}$) of the ellipses. The ellipses are extracted with a minimum of 20 pixels per annulus, and a minimum step size of half FWHM$_\mathrm{beam}$ along the semi-major axis. The inclinations were based on estimates from the ellipses characterising the outer isophotes and global ellipticity respectively of the r-band photometry from the CALIFA survey and are the same as the ones adopted in the stellar dynamical models (section \ref{sect_star}). For most cases, we fix the kinematic position angle ($PA_\mathrm{kin}$) to be the same as $PA_\mathrm{morph}$ (as fitted from the outer isophotes of the r-band photometry). In the few cases where an adjustment of $PA_\mathrm{kin}$ is needed, we allow it to be a free parameter in the extraction of ellipses. Galaxies with adjusted $PA_\mathrm{kin}$ are marked (with *) in Table \ref{tab_galpar}.

The ellipse parameters and rotation velocity of each ellipse are found by fitting a velocity field of the form:
\begin{equation}
V_\mathrm{mod} = V_\mathrm{sys}  + V_\mathrm{rot}\cos(\phi)\sin(i),
\end{equation}
to the observed mean velocity map. Here $V_\mathrm{mod}$, $V_\mathrm{sys}$ and $V_\mathrm{rot}$ are the modelled, systemic and rotation velocities respectively, $\phi$ and $i$ are the azimuthal angle (measured from the major axis) and the inclination respectively. 
To determine the global kinematic centre, $PA_\mathrm{kin}$ and $V_\mathrm{sys}$ of each galaxy, we compute the 
mean of these parameters over all ellipses. The extracted $PA_\mathrm{kin}$ and $V_\mathrm{sys}$ are listed in Table \ref{tab_galpar}. 

To remove any non-circular kinematic perturbations that may come from a bar or spiral arms and could affect our measurement of the rotation curve, we use the method of harmonic decomposition \citep[e.g.][]{kra06,van10}. We model the velocity fields up to their 3rd order harmonics: 
\begin{equation}
\begin{aligned}
&V_\mathrm{mod} = V_\mathrm{sys} + c_{1}\cos(\phi) + s_{1} \sin(\phi) + c_{2}\cos(2\phi) \\
&+ s_{2}\sin(2\phi)+ c_{3}\cos(3\phi) + s_{3}\sin(2\phi),
\end{aligned}
\end{equation}
The obtained value $c_{1}/\sin(i)$ gives us the CO circular velocity (labelled as $V_\mathrm{CO}$ from hereon), largely removing effects from high order perturbations such as for example spiral arms and bars. Whereas the other terms such as $s_{1}$ (radial flow) and the higher order terms are not directly related to the rotation curve, they provide an estimate on the small remaining effects from high order perturbation on $c_{1}$. The relevant higher order terms $c_{1}$, $c_{3}$ and $s_{1}$ are $\lesssim10\%$ of our extracted $V_\mathrm{CO}$. In Appendix \ref{app_c1p} , we estimate the upper limit of the effect of the higher order perturbations in our $V_\mathrm{CO}$ and demonstrate that such perturbations are not correlated with any differences we see between the CO and stellar $V_\mathrm{c}$. 

\subsection{Uncertainty estimates and selection criteria}\label{subsect_uncer}
To estimate the uncertainties in the extracted CO rotation curves, we performed Monte Carlo perturbations of each pixel with a perturbation randomly sampled from a gaussian with width corresponding to the mean velocity error in the corresponding velocity error map. We perform 200 perturbed runs, each time repeating the steps in Section \ref{subsect_vrot}. As our final CO rotation curve, we take the 
mean of the rotation curves extracted from the 200 runs and use that to compare with the CALIFA stellar circular velocities. The standard deviation of the 200 rotation curves is taken as the uncertainty of the rotation curves $\delta V$. We then remove any rotation velocity measurements with $V/\delta V<3$. Finally, we remove the rotation velocity measurements that come from patchy areas in the map as we find that an uneven sampling of line-of-sight velocities along the annuli can render a rotation velocity measurement with large errors (as reflected by a deviation from a smooth rotation curve) that cannot be captured with our estimation of uncertainties. We quantify the patchiness of each annuli by the parameter $P_\mathrm{patch}=\sigma(n_\phi)/\overline{n_\phi}$, where $n_\phi$ is the number of pixels with a velocity measurement per degree in $\phi$ of a particular annuli, $\sigma(n_\phi)$ and $\overline{n_\phi}$ are the standard deviation and mean of $n_\phi$. Rotation velocity measurements from annuli with $P_\mathrm{patch}>1.5$ are removed. After cleaning our sample with the two criteria mentioned above, the average $\delta V$ of our galaxy sample is $\sim$10\,km\,s$^{-1}$. The  extracted $V_\mathrm{CO}$ of UGC04132 are shown in Figure \ref{fig_co_map} as an example. 

\subsection{CO as a kinematically cold tracer}\label{subsect_coadc}
Here, we demonstrate that the CO gas is not pressure supported (i.e. by random motions) and hence our derived rotation curve is a good measure of the circular velocity. Despite being a dynamically cold gas, the CO gas in our sample of galaxies can show a velocity dispersion of up to $\sim$50\,km\,s$^{-1}$ in the inner region. At regions with high velocity dispersion, just like the stellar velocity field, the tangential velocities ($V_\mathrm{\phi}$) can deviate from the true circular velocity ($V_\mathrm{c}$). To estimate this deviation, we applied asymmetric drift corrections (ADC) on the CO rotation curve, which solves the first Jeans equation in the equatorial plane (z=0) such that \citep[rearranged from Eq. A3 of][]{wei08}: 
\begin{equation}
\begin{aligned}
&V_\mathrm{c}^{2}(R)=\overline{V_\mathrm{\phi}}^{2}+\sigma_{R}^{2}\Big[\frac{\partial \mathrm{ln} (\nu\sigma_{R}^{2})}{\partial \mathrm{ln} R}+(\frac{\sigma_\mathrm{\phi}^{2}}{\sigma_{R}^{2}}-1)-\frac{R}{\sigma_{R}^{2}}\frac{\partial \overline{V_{R}V_\mathrm{z}}}{\partial z}\Big], \\
\end{aligned}
\label{eq_adc}
\end{equation}
where $\nu$ is the intrinsic luminosity density, as deprojected from the integrated intensity map of CO, ($V_{R}, V_{z}, V_{\phi}$) and ($\sigma_{R}, \sigma_{z}, \sigma_{\phi}$) are the velocity and velocity dispersion components in the three dimensions of the cylindrical coordinates ($R, z, \phi$). The last term of equation \ref{eq_adc} vanish if we assume the velocity ellipsoid is aligned with the cylindrical coordinate system. Since we do not know how the velocity dispersion is distributed among turbulent, thermal and gravitational dispersions, we take into account the full beam-smeared corrected (see Appendix \ref{app_beamcorr}) velocity dispersion to obtain an upper limit of any possible deviation of the CO $V_\mathrm{\phi}$ to $V_\mathrm{c}$ due to support from random motions. We tested the two limiting cases in which the CO gas is isotropic (i.e. $\sigma_{\phi}^{2}/\sigma_{R}^{2}=1$) and radially anisotropic (i.e. $\sigma_{\phi}^{2}/\sigma_{R}^{2}=0$). In both cases we assume that the CO gas lies on a thin disk with $\sigma_{z}=0$ when deprojecting the velocity dispersion map, such that: $\sigma_{los}^2=\sigma_{R}^2\sin{\phi}^2\sin{i}^2+\sigma_{\phi}^2\cos{\phi}^2\sin{i}^2$. 

\begin{figure}
\includegraphics[trim=280 0 0 320, clip=true,width=0.45\textwidth,angle=90]{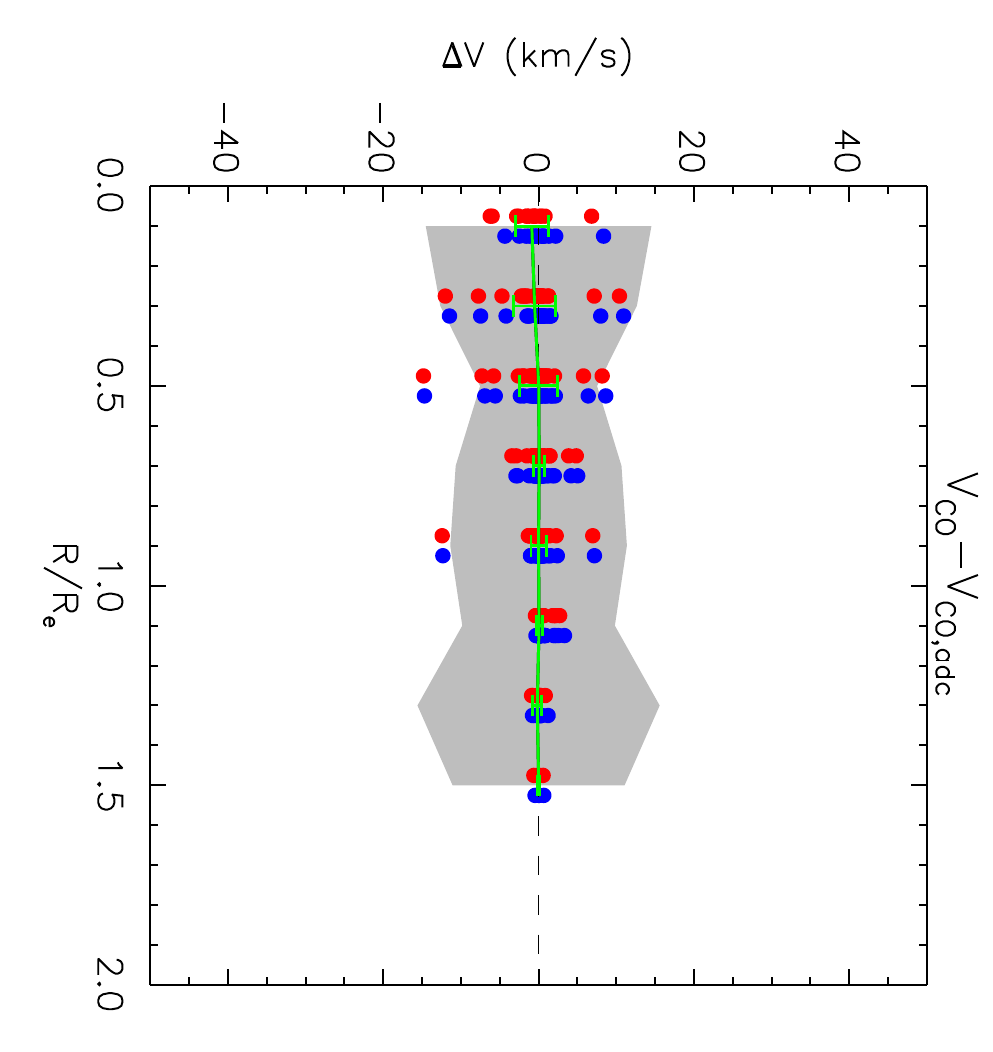}
\caption{We show the differences between the CO rotation curves before and after ADC correction of all the 54 galaxies. Each red/blue dot represents one galaxy in that specific radial bin. The grey slab indicates the average 2$\sigma$ value of the error of all galaxies in our sample in each radius bin. The green curve and error bars indicate the mean and standard deviation of the differences in each bin. As shown in the plot, even in the inner region, the ADC corrections are in fact mostly lying within the  uncertainties.}
\label{fig_pcorr}
\end{figure}

In Figure \ref{fig_pcorr}, we show for all the galaxies in our sample, the difference between the CO rotation curves before and after ADC correction in red for the isotropic case and in blue for the radially anisotropic case, with each dot corresponding to a galaxy at that particular radial bin. We find that the correction to the CO rotation curve is insignificant in either cases, mostly lying even within the error of the rotation curve itself. This suggests that CO is a dynamically cold tracer in our sample of galaxies and the extracted rotation velocity $V_{\phi}$ is a good representation of $V_\mathrm{c}$.

\begin{figure*}
\includegraphics[trim=330 30 0 10, clip=true,width=0.36\textwidth,angle=90.]{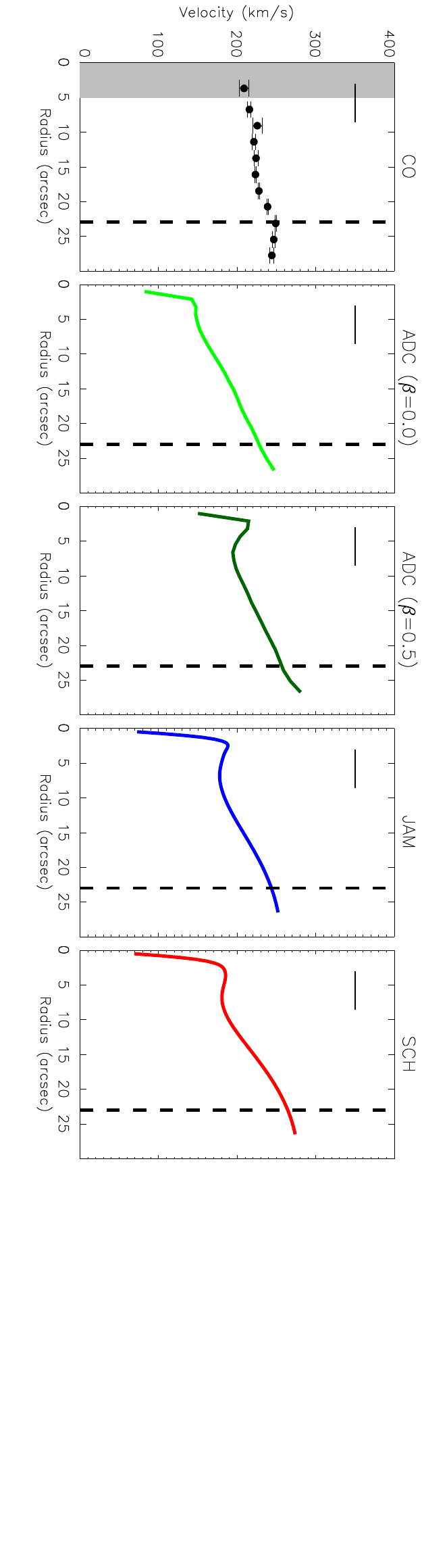}
\caption{We show $V_\mathrm{c}$ of UGC04132 using different kinematic tracers and models. From left to right, the tracer used is: CO gas (black dots, with error bars), and stellar kinematics with the ADC model (green), stellar kinematics with the JAM model (blue) and stellar kinematics with the Schwarzschild model (red). The grey region indicates 3$\sigma_\mathrm{beam}$ of the CO observations. The horizontal line on the top left of each panel indicates the scale of 2\,kpc. The vertical dashed line marks the effective radius. These $V_\mathrm{c}$ from different tracers are stacked on top of each other in Figure \ref{fig_all_vrot} for easier comparison for each galaxy.}
\label{fig_ex_allv}
\end{figure*}

\section{Modelling $V_\mathrm{c}$ from stellar kinematics}\label{sect_star}

We consider three commonly used stellar dynamical models, namely: (1) Asymmetric Drift Correction (ADC), (2) Jeans Anisotropic Models (JAM) and (3) Schwarzschild models (SCH). As mentioned in the Introduction, out of these three models, ADC is the most easily implemented and requires the largest amount of assumptions. SCH, on the other hand, require the fewest assumptions but is the most computationally expensive method. Below we outline the methods and assumptions behind each of the models, which we summarise in Table \ref{tab_prop}.

\begin{table}
\begin{center}
\begin{tabular}{|c|c|c|c|}
\hline
\makecell{Method} & ADC & JAM & Schwarzschild \\
\cline{2-4}
 & \multicolumn{2}{c|}{Solving Jeans Equations} & Orbit-based\\
\hline
\makecell{Geometric \\Assumption} & \makecell{Axisymmetric\\Thin Disk} & \makecell{Axisymmetric \\3D} & Triaxial \\
\hline
\makecell{Constant \\ M/L} & X & \multicolumn{2}{c|}{Luminous matter} \\
\hline
\makecell{Dark matter\\ halo} & X & \multicolumn{2}{c|}{Spherical NFW}\\
\hline
\makecell{Velocity\\Ellipsoid} & \multicolumn{1}{c|}{\makecell{$\beta=0.0$ or \\$\beta=0.5$}} &\multicolumn{1}{c|}{\makecell{Constant \\Anisotropy}} & \makecell{No \\Assumption}\\
\cline{2-3}
& \multicolumn{2}{c|}{Aligned with cylindrical coord.} &\\
\hline

\end{tabular}
\end{center}
\caption{Properties and assumptions of the three stellar dynamical models: ADC, JAM and Schwarzschild models. "X" indicates that the respective parameter is not incorporated in that specific model. \label{tab_prop}}
\end{table}

\subsection{Asymmetric Drift Correction (ADC)}

As described in section \ref{subsect_coadc}, ADC solve the Jeans equations utilising the line-of-sight mean velocity and velocity dispersion maps, adopting a thin disk assumption by solving the Jeans equation only in the $z=0$ plane (i.e. equation \ref{eq_adc}), and in addition assumes an axisymmetric gravitational potential. In solving equation \ref{eq_adc}, we assume that the velocity ellipsoid aligns with cylindrical coordinates and that the velocity anisotropy is constant. We derive $V_\mathrm{ADC}$ for all the galaxies in our sample with two commonly assumed values of the velocity anisotropy $\beta=1-\sigma_\phi^{2}/\sigma_{r}^{2}$: $\beta=0.0$ (isotropic) and $\beta=0.5$ (radially anisotropic) \citep[e.g.][]{hinz01,lea12}. To derive smooth surface brightness profiles $\nu$, we fitted Multi Gaussian Expansions (MGEs) \citep{em94} to SDSS r-band images. We also fitted a power law to the extracted $V_\phi$ and an exponential profile to $\sigma_{R}$ to ensure a smooth $V_\mathrm{c}$. The functional form of the fittings are:
\begin{equation}
\begin{aligned}
&V_\phi=V_{0}\frac{R}{(R_{c}^{2}+R^{2})^{0.5+0.25\alpha}}\\
&\sigma_R=\sigma_{0}e^{-R/R_{s}}+\sigma_\infty,
\end{aligned}
\end{equation}
where ($V_{0}, R_{c}, \alpha$) and ($\sigma_0, R_s, \sigma_\infty$) are the free parameters in the fitting of $V_\phi$ and $\sigma_{R}$ respectively. The fitted MGEs, the extracted and fitted $V_\phi$ and $\sigma_{R}$ (for the case of $\beta=0.5$) of all the galaxies in our sample can be found in Appendix \ref{app_stellarmod}, those of UGC04132 are shown here in Figure \ref{fig_UGC04132star} as an example. The circular velocities extracted using ADC are labelled as $V_\mathrm{ADC}$ in the rest of the paper, the two specific cases with $\beta=0.0$ and $\beta=0.5$ are labelled as $V_\mathrm{ADC, \beta=0.0}$ and $V_\mathrm{ADC, \beta=0.5}$. In Figure \ref{fig_ex_allv}, we show $V_\mathrm{ADC, \beta=0.0}$ and $V_\mathrm{ADC, \beta=0.5}$ for UGC04132 in light and dark green curves respectively.

\subsection{Axisymmetric Jeans Anisotropic Multi-Gaussian Expansion Models (JAM)}\label{subsect_jam}
JAM also solves the Jeans equations utilising the line-of-sight mean velocity and velocity dispersion maps, but under different assumptions. Just like with ADC, JAM assumes an axisymmetric gravitational potential and a velocity ellipsoid aligned with the cylindrical coordinate system. Unlike ADC however, JAM takes into account a full line-of-sight integration when modelling the observed velocity moments. It involves two of the Jeans equations (all the terms in the third Jeans equation vanish due to the axisymmetric assumption): 
\begin{equation}
\begin{aligned}
&\frac{\partial(R\nu\overline{V_{R}^{2}})}{\partial R}+R\frac{\partial(\nu\overline{V_{R}V_\mathrm{z}})}{\partial z}-\nu\overline{V_\mathrm{\phi}^{2}}+R\nu\frac{\partial \Phi}{\partial R}=0,\\
&\frac{\partial(R\nu\overline{V_{R}V_\mathrm{z}})}{\partial R}+R\frac{\partial(\nu\overline{V_\mathrm{z}^{2}})}{\partial z}+R\nu\frac{\partial \Phi}{\partial z}=0,
\end{aligned}
\end{equation}
where $\nu(R,z)$ is the intrinsic luminosity density and $\Phi(R,z)$ is the axisymmetric gravitational potential. Again, ($V_{R}, V_{z}, V_{\phi}$) are the velocity components in the three dimensions of the cylindrical coordinates ($R, z, \phi$). 

We use the JAM code developed by \cite{cap08}\footnote{we use the python version of code which can be downloaded from http://www-astro.physics.ox.ac.uk/$\sim$mxc/software} to construct the modelled kinematics. In our models, the gravitational potential is composed of two components: a luminous component and a dark matter halo. For the luminous component, we follow the commonly adopted mass-follow-light assumption. We again describe the light distribution $\nu(R,z)$ with the same MGEs as used in our ADC, and multiply that with a constant stellar mass-to-light ratio $\Upsilon_\star$ to obtain the mass distribution of the luminous matter, which we assume to be axisymmetric. A spherical NFW \citep{nfw96} dark matter halo is then added to the potential, with the concentration fixed to be related to the virial mass $M_{200}$ \citep{dut14}, defined as the enclosed mass within $r_{200}$. In addition, we assume a constant velocity anisotropy $\beta_{z}=1-\sigma_{z}^{2}/\sigma_{r}^{2}$ in our JAM models. There are hence in total three free parameters in the fitting of the models: the stellar mass-to-light ratio $\Upsilon_\star$, the virial velocity of the dark matter halo $V_\mathrm{vir}$, and the velocity anisotropy $\beta_{z}$. The modelled kinematics are then fitted to the observed kinematics via the term $V_\mathrm{rms}=\sqrt{V_\mathrm{los}^2+\sigma_\mathrm{los}^{2}}$, where $V_\mathrm{los}$ is the line-of-sight mean velocity and $\sigma_\mathrm{los}$ is the line-of-sight projected velocity dispersion.  

\begin{figure}
\includegraphics[trim=20 0 0 0,width=0.5\textwidth]{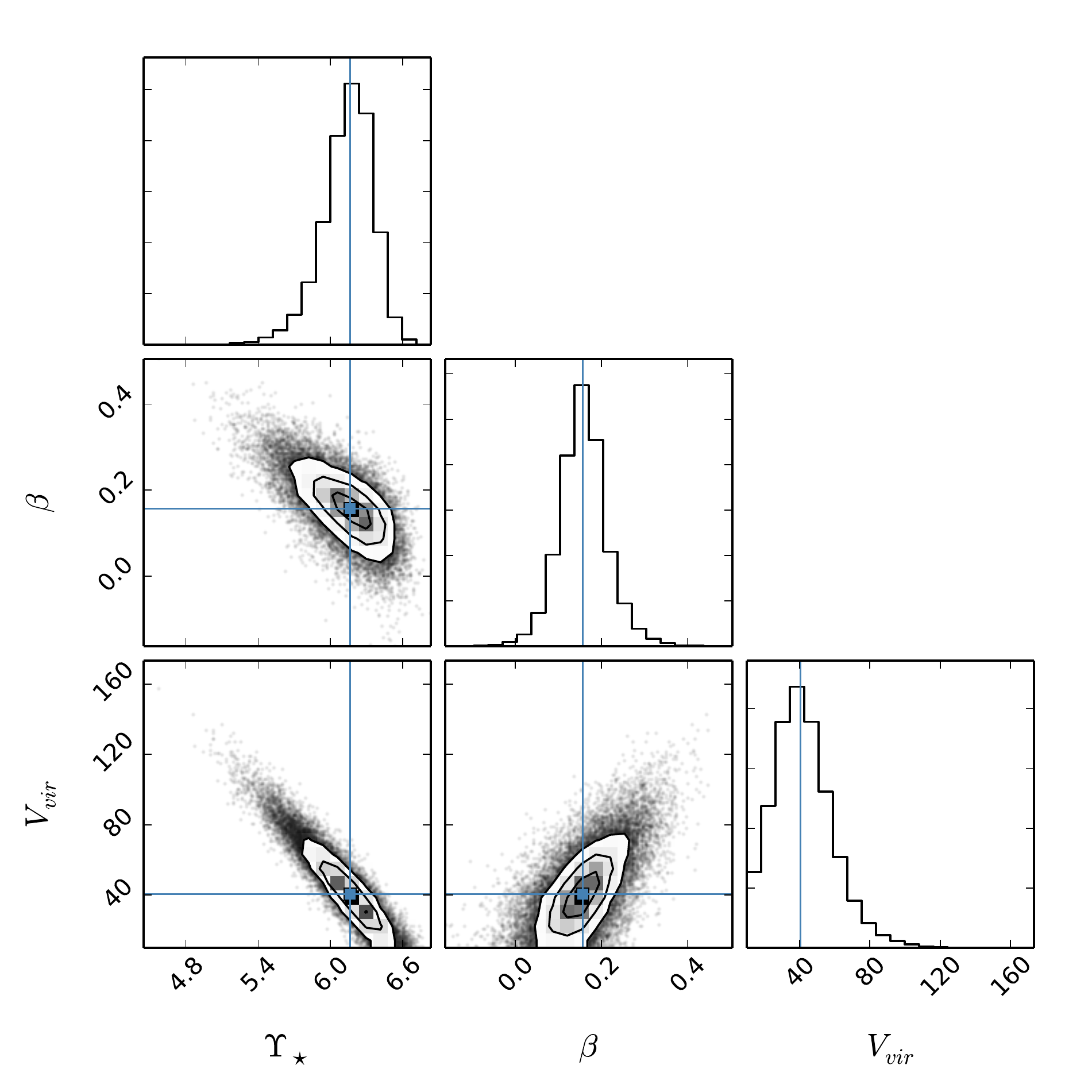}
\caption{Posterior and covariance distributions return from our MCMC analysis for parameters in the JAM model of UGC04132. Contours show the 1, 2 and 3$\sigma$ constrains on each parameter.}
\label{fig_emcee}
\end{figure}

\begin{figure*}
\begin{center}
\includegraphics[trim=270 50 120 50,width=0.3\textwidth,angle=90]{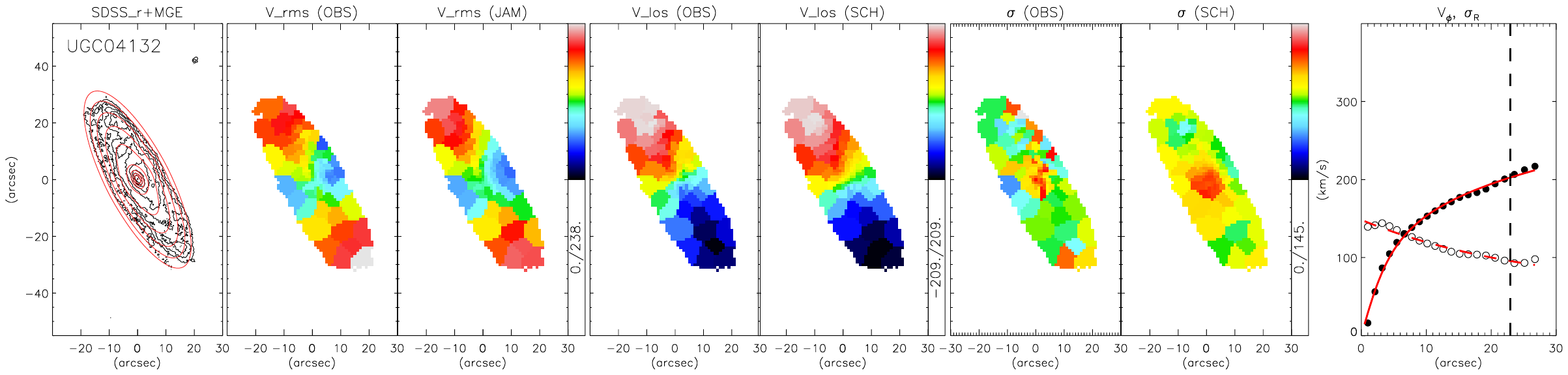}
\caption{Observables and best fit models of stellar dynamical models of UGC04132. From left to right: (1) r-band image from SDSS plotted in black contours with the fitted MGEs are over-plotted with red contours, (2) observed $V_\mathrm{rms}$ from the CALIFA survey, (3) best fitted JAM modelled $V_\mathrm{rms}$, (4)$V_\mathrm{los}$ from the CALIFA survey, (5) $V_\mathrm{los}$ from the Schwarzschild model , (6) observed $\sigma_\mathrm{los}$, (7) modelled $\sigma_\mathrm{los}$, (8) extracted $V_\phi$ and $\sigma_R$ (for the case of $\beta=0.5$) values in solid and open circles, and the fitted functional forms in solid and dashed red lines respectively.}
\label{fig_UGC04132star}
\end{center}
\end{figure*}

We constrain the fitting of the kinematics by the Markov-Chain Monte-Carlo method (MCMC), implemented with the publicly available software $\tt{emcee}$\footnote{the software can be found on https://github.com/dfm/emcee} \citep{emcee}. We employ 100 walkers and 500 steps when modelling each of the galaxies, with a burn-in phase of 50 steps. We apply uniform priors of $0.5<\Upsilon_\star<10$, $0\,\mathrm{km/s}<V_\mathrm{vir}<400\,\mathrm{km/s}$ and $-2<\beta_{z}<1$. We assume that the observation errors are gaussian and adopt $\mathscr{L}=\exp\big(-\frac{\chi^2}{2}\big)$ as our likelihood function. For most galaxies, the free parameters converge well within our imposed priors. We show in Figure \ref{fig_emcee}, the posterior distribution of the parameter space for galaxy UGC04132 as a representative example. The observed and modelled $V_\mathrm{rms}$ of this particular galaxy is shown in Figure \ref{fig_UGC04132star}. The MGE and $V_\mathrm{rms}$ fittings for the rest of the galaxies in our sample are shown in Appendix \ref{app_stellarmod}. We label the circular velocities extracted using JAM as $V_\mathrm{JAM}$ from hereon. $V_\mathrm{JAM}$ of UGC04132 is shown in blue in Figure \ref{fig_ex_allv}. 

The best fitted parameters for all the galaxies, and the reduced $\chi^{2}$ of our best fit models are listed in Table \ref{tab_jam}. We note that for 7 galaxies in our sample, $\beta_{z}$ is driven to the lower limit of our imposed prior, as marked with $\dagger$ in Table \ref{tab_jam}. Such behaviours persist even if we allow the inclination to vary, as we show in Appendix \ref{app_jam}, suggesting the behaviours are intrinsic to the JAM models for these galaxies and do not arise from incorrect assumptions of inclinations. Additionally, for 7 galaxies in our sample, $V_\mathrm{vir}$ is driven to the upper limit of our imposed prior, as marked with $\ddagger$ in Table \ref{tab_jam}. To improve the fits for these galaxies, we further impose constraints from studies of abundance matching in simulations and empirical stellar-halo mass relations. We adopt the function form outlined in \cite{leau12}: 
\begin{equation}
\log(M_h)=\log(M_1)+\beta\log\Big(\frac{M_s}{M_{0}}\Big)+\frac{(\frac{M_s}{M_{0}})^\delta}{1+(\frac{M_s}{M_{0}})^{-\gamma}}-0.5,
\label{eq_smhm}
\end{equation}
where $M_h$ is the halo virial mass and $M_s$ is the total stellar mass, which is the integrated mass from the MGEs multiplied by $\Upsilon_\star$. We adopt the parameters $\beta=0.456$, $\delta=0.583$, $\gamma=1.48$, $\log(M_0)=10.917$ and $\log(M_1)=12.518$ from \cite{leau12}. We further discuss both issues and how they might affect our results in Appendix \ref{app_jam}.

\begin{table}
\begin{threeparttable}
\begin{tabular}{@{}lllll}
\hline
Galaxy& $\Upsilon_\star$ & $\beta_{z}$ & $V_\mathrm{vir}$\,(km/s) & reduced $\chi^2$\\
\hline
\hline
IC0480&    5.66&    0.73&     59.19&    2.86\\
IC0944&    5.90&    0.43&    181.17&    3.19\\
IC1199&    5.54&    0.54&    132.25&    1.53\\
IC1683&    3.39&   -1.41&    223.03&    1.45\\
IC2247&    5.15&    0.67&     98.38&    1.87\\
IC2487&    7.52&    0.07&     48.88&    1.17\\
IC4566&    5.04&   -0.73&     57.20&    2.39\\
NGC0477$^{\dagger}$&    4.99&   -1.95&     44.88&    1.50\\
NGC0496&    2.43&    0.40&     67.53&    0.94\\
NGC0551&    4.32&    0.23&     34.36&    1.98\\
NGC2253$^{\dagger}$&    2.32&   -1.97&     98.41&    1.26\\
NGC2347$^{\ddagger}$&    3.89&   -0.74&    109.52&    2.24\\
NGC2410&    5.68&    0.23&     18.25&    1.28\\
NGC2639$^{\ddagger}$&    3.64&    0.82&    111.16&    4.22\\
NGC2906&    4.49&    0.13&    193.42&    0.91\\
NGC3815&    3.57&    0.55&    122.62&    1.74\\
NGC3994&    3.49&    0.47&     14.53&    3.46\\
NGC4047$^{\dagger}$&    2.58&   -1.94&    190.30&    0.73\\
NGC4149&    5.40&    0.44&     24.23&    8.28\\
NGC4210$^{\dagger}$&    4.11&   -1.94&     32.17&    2.52\\
NGC4644$^{\dagger}$&    4.50&   -1.90&     96.82&    2.29\\
NGC4711&    3.66&   -0.15&     47.95&    0.83\\
NGC4961$^{\ddagger}$&    2.92&    0.40&     78.35&    1.75\\
NGC5016&    3.15&   -0.57&    131.73&    0.58\\
NGC5056$^{\dagger}$&    4.09&   -1.97&     19.85&    2.57\\
NGC5218$^{\ddagger}$&    7.20&    0.61&    104.53&    1.81\\
NGC5480$^{\dagger}$&    2.44&   -1.90&     19.39&    1.32\\
NGC5520&    3.74&   -0.01&    351.46&    1.39\\
NGC5633&    3.24&    0.47&     50.70&    1.57\\
NGC5784$^{\ddagger}$&    3.84&   -0.54&    121.16&    7.51\\
NGC5908$^{\ddagger}$&    5.57&    0.51&    111.33&   10.29\\
NGC5980&    4.50&    0.56&     15.65&    5.45\\
NGC6060&    4.97&    0.31&     32.34&    0.40\\
NGC6168&    2.76&    0.88&    113.70&    0.74\\
NGC6186&    3.53&    0.61&     14.40&    5.36\\
NGC6301&    4.92&   -0.86&    106.80&    3.98\\
NGC6394&    5.14&   -0.02&    114.90&    4.74\\
NGC6478&    4.48&    0.10&     94.09&    4.82\\
UGC00809&    8.37&    0.54&     90.07&    2.33\\
UGC03539&    5.02&    0.89&    255.40&    1.88\\
UGC03969&    6.36&    0.77&     71.45&    1.98\\
UGC04029&    6.14&    0.65&     61.95&    7.50\\
UGC04132&    6.16&    0.16&     40.48&    2.40\\
UGC05108&    5.89&   -0.52&    288.76&    1.89\\
UGC05598&    4.02&    0.73&     75.32&    1.16\\
UGC08107&    9.91&    0.41&    175.84&    1.80\\
UGC09067&    4.52&    0.32&     33.67&    1.78\\
UGC09537$^{\ddagger}$&    6.28&    0.10&    134.90&    5.64\\
UGC09542&    4.05&    0.79&     83.43&    1.87\\
UGC09665&    2.74&    0.82&    315.18&    0.71\\
UGC09892&    3.40&    0.12&     48.85&    0.66\\
UGC10123&    5.19&    0.65&    340.85&    2.25\\
UGC10384&    3.82&    0.81&    255.11&    1.94\\
UGC10710&    4.80&    0.47&    148.44&    2.86\\
\hline
\end{tabular}
\caption{Best fitted parameters and reduced $\chi^{2}$ of our JAM models. $\dagger$ marks the galaxies which have best fitted $\beta_\mathrm{z}<-1.5$, and $\ddagger$ marks the galaxies for which we impose an additional stellar-mass-halo-mass relation from \protect\cite{leau12}. }
\label{tab_jam}
\end{threeparttable}
\end{table}

\subsection{Schwarzschild Models (SCH)}
The Schwarzschild models adopt a different approach. Instead of solving the Jeans equations, the Schwarzschild models compute the orbits in a gravitational potential to recover the observed kinematics. A complete description to the methodology of our Schwarzschild models can be found in \cite{zhu18} and the resulting orbital distribution derived for the CALIFA galaxies and their fitted parameters as adopted here can be found in \cite{zhu17}. Here we give a brief overview of our Schwarzschild models for completeness. First, a set of mock triaxial gravitational potentials are created. Each of the gravitational potentials is described by two components: mass from luminous matter and mass from dark matter. The stellar mass-to-light ratio is assumed to be constant: $\Upsilon_\star$, with the light distribution again modelled with MGEs. Unlike in JAM however, the luminous mass distributions in our Schwarzschild models are allowed to be triaxial. The triaxial luminous mass distributions are characterised by the two parameters $p$ and $q$, which are the ratio between the intermediate axis and short axis with the long axis respectively. Again, the dark matter component is assumed to follow a spherical NFW profile, with the same mass-concentration relation as adopted in JAM. The free parameters here therefore include only the stellar mass-to-light ratio $\Upsilon_\star$, the virial mass $M_{200}$ and the triaxial parameters $(p,q)$. For each of the mock potentials, an orbit library is computed. The orbits in the library are then weighted and used to create mock line-of-sight mean velocity and velocity dispersion maps. The mock kinematic maps (both $V_\mathrm{los}$ and $\sigma_\mathrm{los}$) are then fitted to the observed kinematic maps to constrain the weight of each orbit. The gravitational potential with which its best-fitted orbital weights provide the best fit to the observed map is chosen as the best estimate of the true gravitational potential. Finally, the circular velocity is calculated from this best-fit gravitational potential. The Schwarzschild model therefore does not put assumptions on the velocity ellipsoid but still assumes a constant stellar mass-to-light ratio and an NFW profile for the dark matter halo. To allow the readers an assessment to how well the Schwarzschild models are fitted to the kinematics, we include the observed and best fitted Schwarzschild model kinematics of our full sample of galaxies in Appendix \ref{app_stellarmod} and show here in Figure \ref{fig_UGC04132star}, those of UGC04132 as an example. We label the circular velocities extracted from the Schwarzschild models as $V_\mathrm{SCH}$. $V_\mathrm{SCH}$ of UGC04132 is shown in red in Figure \ref{fig_ex_allv}.

\begin{figure*}
\centering
\includegraphics[trim=30 40 60 50, clip=true, page=1, width=0.65\textwidth,angle=90]{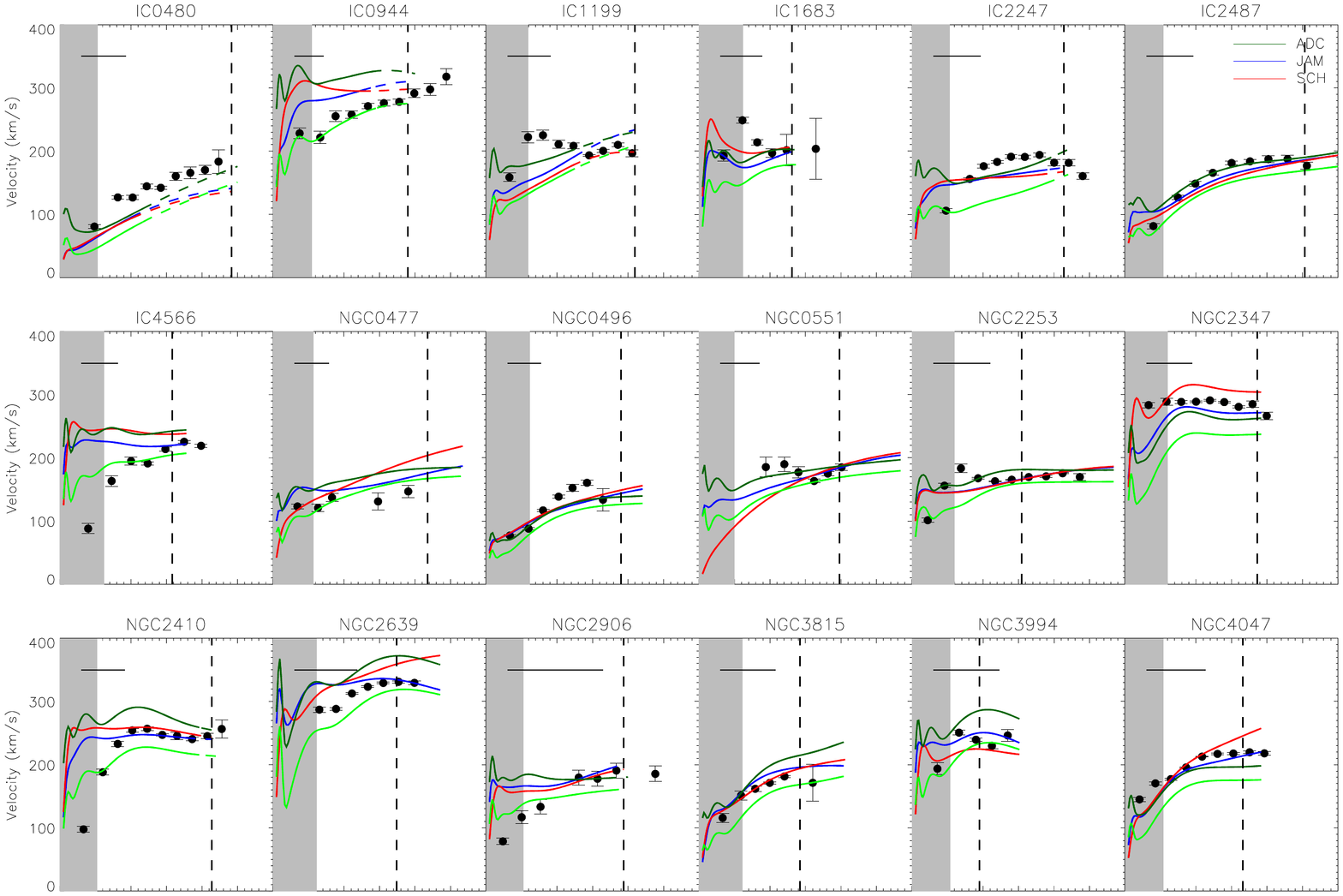}
\includegraphics[trim=30 40 60 50, clip=true, page=2, width=0.65\textwidth,angle=90]{all_vc.pdf}
\end{figure*}
\begin{figure*}
\centering
\includegraphics[trim=30 40 60 50, clip=true, page=3,width=0.65\textwidth,angle=90]{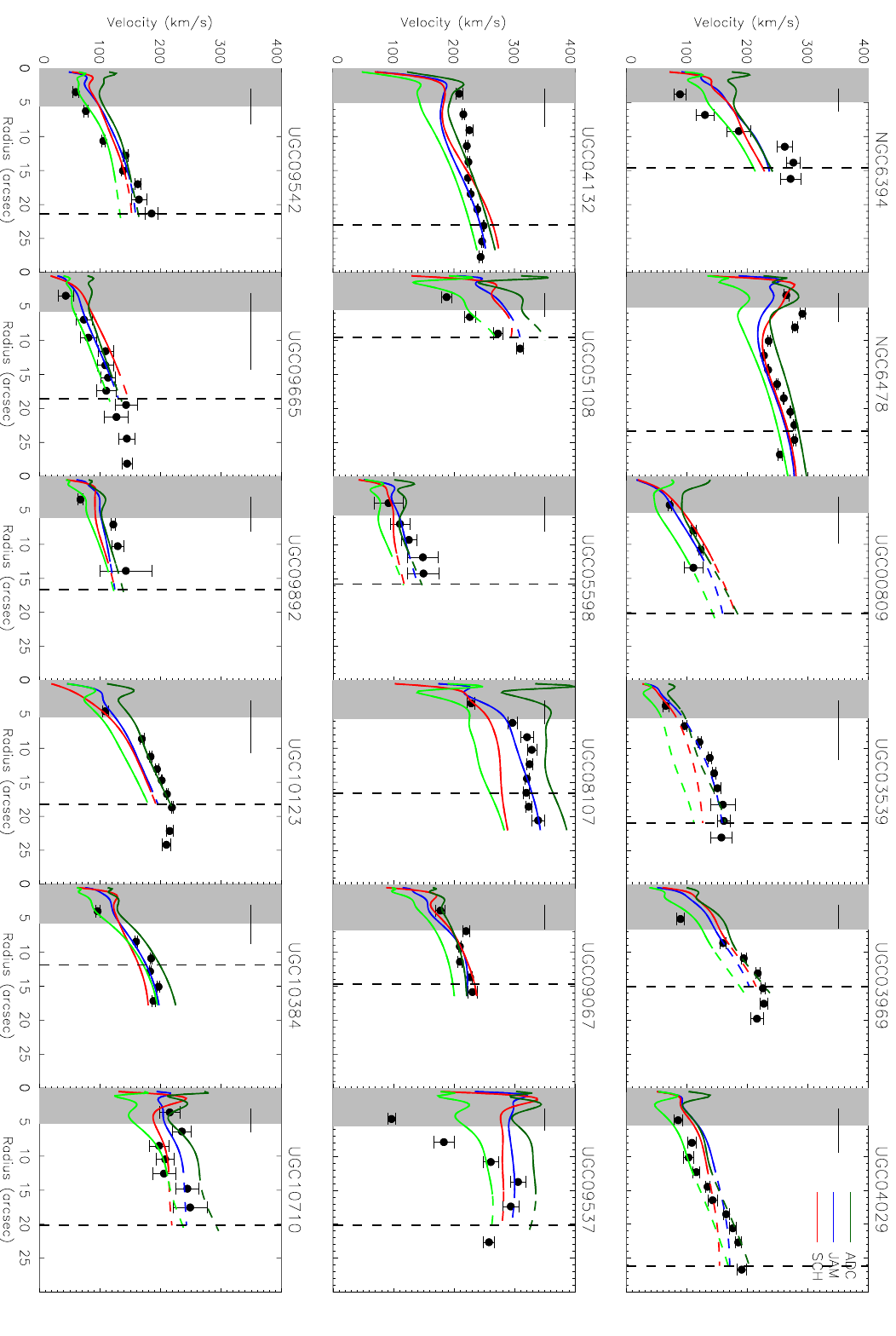}
\caption{Circular velocities of the 54 galaxies. $V_\mathrm{CO}$ obtained in this work are plotted in black dots, with the error indicated by the error bars. $V_\mathrm{ADC, \beta=0.0}$, $V_\mathrm{ADC, \beta=0.5}$, $V_\mathrm{JAM}$ and $V_\mathrm{SCH}$ are plotted in light green, dark green, blue and red curves respectively. The horizontal line on the top left of each panel indicate the scale of 2\,kpc. The vertical dashed line marks the effective radius. The grey region indicate 3$\sigma_\mathrm{beam}$ of the CO observations. 
}
\label{fig_all_vrot}
\end{figure*}

\section{Differences of $V_\mathrm{c}$ extracted from CO and stellar kinematics}\label{sect_costar}
In this section we describe the comparison of $V_{c}$ extracted using different kinematic tracers: dynamically cold molecular tracer CO and dynamically hot stellar kinematics, including those derived from the Asymmetric Drift Correction (ADC), Jeans (JAM) and Schwarzschild (SCH) models. All the $V_{c}$ for our sample of 54 galaxies extracted with the aforementioned kinematic tracers are presented in Figure \ref{fig_all_vrot}. We first compare the different stellar dynamical models with CO in the following order: ADC vs. CO, JAM vs. CO and SCH vs. CO. For each model, we begin by comparing the stellar and CO $V_\mathrm{c}$ at one effective radius, and then we characterise the variation of the differences with respect to galactic radii, stellar $V_\phi/\sigma_{R\star}$ values and galactic properties. We then also examine how the three stellar dynamical models perform when compared against each other.

\begin{figure*}
\begin{center}
\includegraphics[trim=30 70 30 120, clip=true,width=1.0\textwidth]{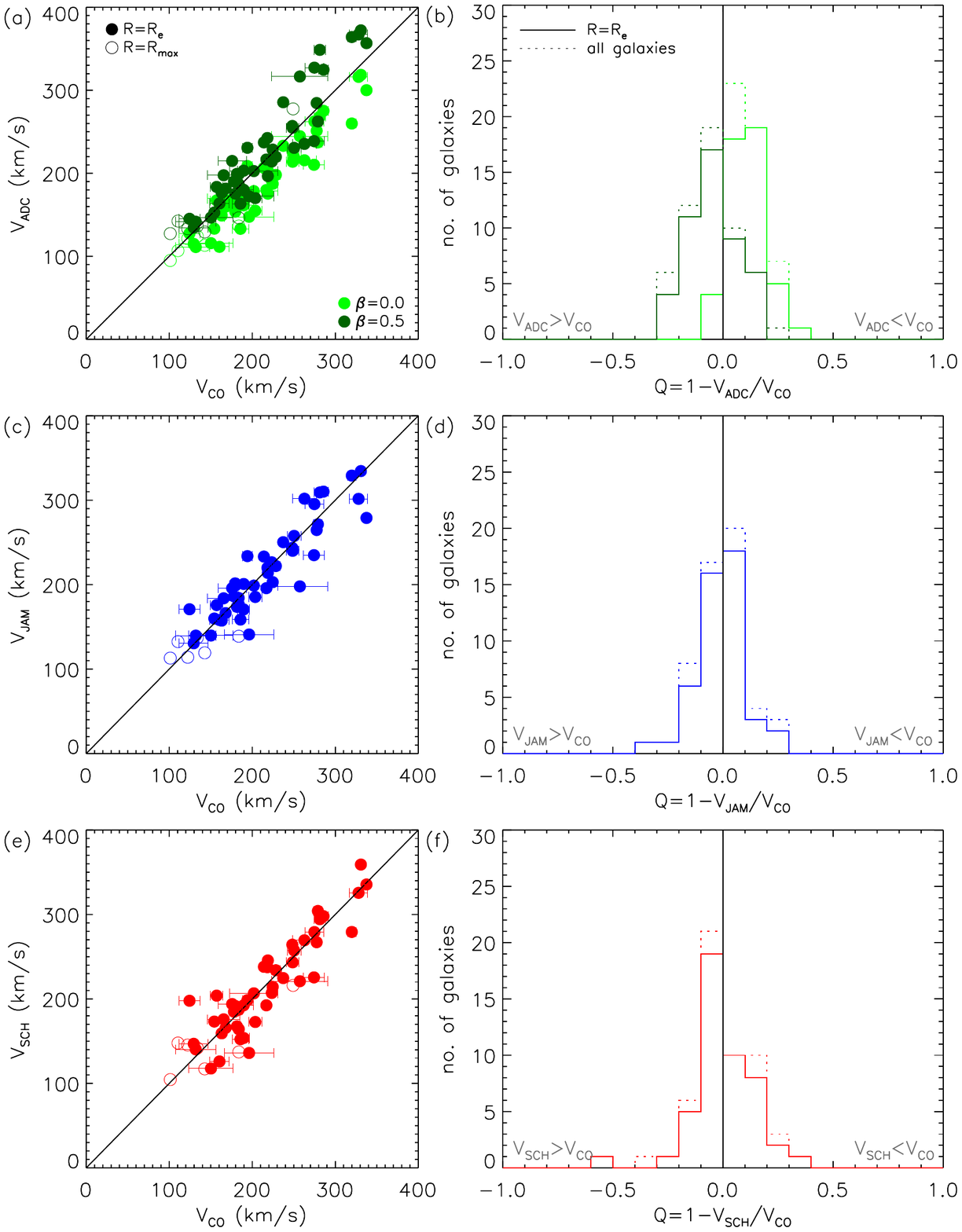}
\caption{Comparison between the stellar and CO circular velocities at 1 $R_\mathrm{e}$. Panels (a), (c) and (e) show $V_\mathrm{CO}$ plotted against $V_\mathrm{ADC}$, $V_\mathrm{JAM}$ and $V_\mathrm{SCH}$ respectively, with the black line indicating the one-to-one line. It is shown here that $V_\mathrm{ADC}$ underestimate the circular velocity with $\beta=0.0$, but agree well with $V_\mathrm{CO}$ with $\beta=0.5$, except for high-mass galaxies. Also, both $V_\mathrm{JAM}$ and $V_\mathrm{SCH}$ agree well with $V_\mathrm{CO}$ at $R_\mathrm{e}$. Panels (b), (d) and (f) show the relative difference, $Q_{X}$, for ADC, JAM and SCH respectively. The black vertical lines indicate $Q=0$, to the right of the black lines are galaxies from which the stellar $V_\mathrm{c}$ is smaller than $V_\mathrm{CO}$, again a bias is seen for $V_\mathrm{ADC, \beta=0.0}$, but none in $V_\mathrm{ADC, \beta=0.5}$, $V_\mathrm{JAM}$ and $V_\mathrm{SCH}$.}
\label{fig_reff}
\end{center}
\end{figure*}

\begin{figure*}
\begin{center}
\includegraphics[trim=180 20 215 20, clip=true,width=0.32\textwidth,page=1,angle=90]{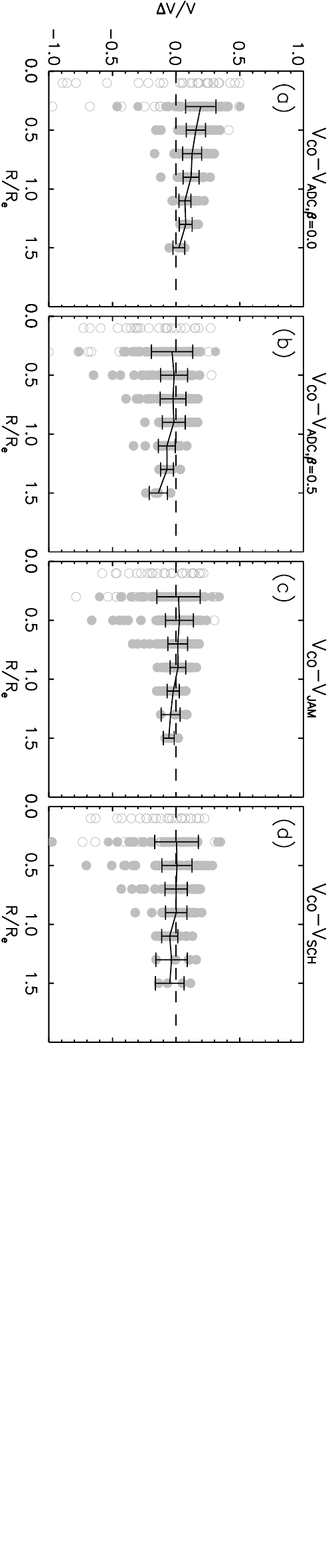}
\caption{Velocity differences between the stellar and CO circular velocity curves in radial bins. Each grey dot represent a measurement from one galaxy at that specific radial bin. The error-weighted mean and standard deviation of each bin are shown in black curve and error bars respectively. $V_\mathrm{ADC, beta=0.0}$ underestimate $V_\mathrm{c}$ at all radii, with increasing disagreement with the intrinsic value towards the inner region. While on average, $V_\mathrm{ADC, beta=0.5}$, $V_\mathrm{JAM}$ and $V_\mathrm{SCH}$ agree with CO at all radii, a large scatter can be seen in the inner region. The open grey circles indicate measurements at $R<3\sigma_\mathrm{beam}$, the corresponding mean and standard deviation are marked with dotted lines.}
\label{fig_global_r}
\end{center}
\end{figure*}

\begin{figure*}
\begin{center}
\includegraphics[trim=180 20 215 20, clip=true,width=0.32\textwidth,page=1,angle=90]{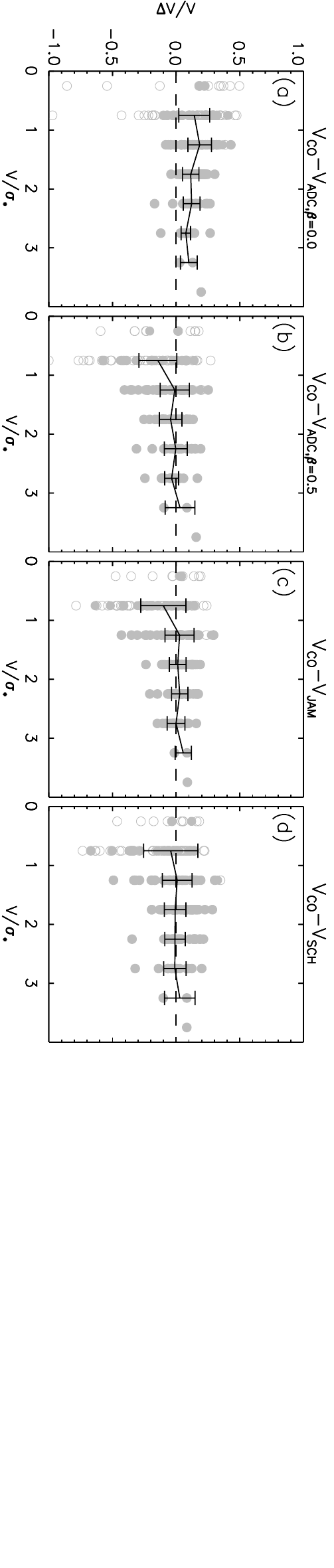}
\caption{Velocity differences between the stellar and CO circular velocity curves in $V/\sigma_\star$ bins. Each grey dot represent a measurement from one galaxy at that specific $V/\sigma_\star$ bin. The error-weighted mean and standard deviation of each bin are shown in black curve and error bars respectively. Again, $V_\mathrm{ADC, beta=0.0}$ underestimate $V_\mathrm{c}$ at all $V/\sigma_\star$ bins. All $V_\mathrm{ADC, beta=0.5}$, $V_\mathrm{JAM}$ and $V_\mathrm{SCH}$ agree well with CO on average in all $V/\sigma_\star$ bins. A large scatter can be seen towards the low $V/\sigma_\star$ regime. The open grey circles indicate measurements at $R<3\sigma_\mathrm{beam}$.}
\label{fig_global_vod}
\end{center}
\end{figure*}

\subsection{ADC vs. CO}
In Figure \ref{fig_reff}(a), we plot in solid circles the values of $V_\mathrm{ADC}$ versus $V_\mathrm{CO}$ at the effective radii $R_\mathrm{e}$ for the 47 galaxies in our sample in which $V_\mathrm{CO}$ reaches 1\,$R_\mathrm{e}$. For galaxies where the observed CO kinematics reaches 1\,$R_{e}$ while the observed stellar kinematics do not, we extrapolate $V_\mathrm{ADC}$ with the MGEs, the fitted power-law for $V_\phi$ and the fitted exponential law for $\sigma_R$. The extrapolated $V_\mathrm{ADC}$ are shown as dashed lines in Figure \ref{fig_all_vrot}. In open circles, we plot $V_\mathrm{ADC}$ versus $V_\mathrm{CO}$ at the maximum observed radius ($R_\mathrm{max}$) for the remaining 9 galaxies for references. We do not extrapolate $V_\mathrm{CO}$.

Light green circles denote $V_\mathrm{ADC, \beta=0.0}$ and dark green circles denote $V_\mathrm{ADC, \beta=0.5}$. This plot indicates visually that $V_\mathrm{ADC, \beta=0.0}$ is smaller than $V_\mathrm{CO}$ at $R_\mathrm{e}$ in general. On the other hand, $V_\mathrm{ADC, \beta=0.5}$ mostly agree well with $V_\mathrm{CO}$, with the exception of the few highest mass galaxies with $V_\mathrm{CO}\gtrsim280$\,km\,s$^{-1}$. $V_\mathrm{ADC, \beta=0.5}$ tend to overestimate $V_\mathrm{c}$ on the high-mass end at $R_\mathrm{e}$. To quantify any biases or agreements, we compute the relative difference $Q_\mathrm{ADC}=(1-\frac{V_\mathrm{ADC}}{V_\mathrm{CO}})_{R_\mathrm{e}}$. The histogram of $Q_\mathrm{ADC}$ is shown in Figure \ref{fig_reff}(b) in solid lines for galaxies with $R_\mathrm{max}>R_\mathrm{e}$, and in dashed line we show the histogram for all galaxies, with $Q$ being computed at $R=R_\mathrm{max}$ for galaxies which have $R_\mathrm{max}<R_\mathrm{e}$. Considering only the galaxies which are observed beyond 1\,$R_\mathrm{e}$, the mean and standard deviation of $Q_\mathrm{ADC, \beta=0.0}$ are 11\% and 6\% respectively, confirming $V_\mathrm{ADC, \beta=0.0}$ is smaller than $V_\mathrm{CO}$ on average. $V_\mathrm{ADC, \beta=0.5}$ shows a better agreement with $V_\mathrm{CO}$, with the mean and standard deviation of $Q_\mathrm{ADC, \beta=0.5}$ being $-$5\% and 8\% respectively.

We next investigate how the difference $\Delta V_\mathrm{ADC}$ ($=V_\mathrm{CO}-V_\mathrm{ADC}$) varies with galactic radii. In Figure \ref{fig_global_r}(a) and (b), we show the relative difference $\Delta V_\mathrm{ADC}/V_\mathrm{CO}$ for $\beta=0.0$ and $\beta=0.5$ respectively, plotted against normalised radii $R/R_\mathrm{e}$. Circular velocities of each galaxy are first binned into radial bins as listed in Table \ref{tab_r}. Then we compute a value for $\Delta V_\mathrm{ADC}/V_\mathrm{CO}$ for each bin in each galaxy, corresponding to a grey point in Figure \ref{fig_global_r}(a). Then for each radial bin, we compute the average and standard deviation over all galaxies, shown as the black curve and error bars in Figure \ref{fig_global_r}(a), with values listed in Table \ref{tab_r}. We shall restrict our discussion to bins that are outside 3$\sigma$ of the radio beam ($\sigma_\mathrm{beam}$); even though we already performed a beam smearing correction, an uneven distribution of CO gas within the beam can still affect the resulting $V_\mathrm{CO}$. We still show the bins within 3$\sigma_\mathrm{beam}$ for reference in Figure \ref{fig_global_r} with open circles. 

Figure \ref{fig_global_r}(a) shows an increasing trend in mean $\Delta V_\mathrm{ADC, \beta=0.0}/V_\mathrm{CO}$ towards the center, indicating that the isotropic ADC increasingly underestimate $V_\mathrm{c}$ towards the central regions of galaxies. Within 1$R_\mathrm{e}$, $V_\mathrm{ADC, \beta=0.0}$ underestimate $V_\mathrm{c}$ by $\sim$13\% on average, with the scatter of $\Delta V_\mathrm{ADC, \beta=0.0}/V_\mathrm{CO}$ increasing towards the center to $\sim$12\%. On the other hand, $V_\mathrm{ADC, \beta=0.5}$ perform better than $V_\mathrm{ADC, \beta=0.0}$ in all radial bins with $R<R_\mathrm{e}$, as shown in Figure \ref{fig_global_r}(b). At $R<R_\mathrm{e}$, $V_\mathrm{ADC, \beta=0.5}$ and $V_\mathrm{CO}$ agree to within 1\,$\sigma$. Just like the case with $\beta=0.0$, the scatter in $\Delta V_\mathrm{ADC, \beta=0.5}/V_\mathrm{CO}$ increases towards the inner region to $\sim$16\%. 

We show a similar plot of $\Delta V_\mathrm{ADC}/V_\mathrm{CO}$, but against $V_{\phi}/\sigma_{R\star}$, in Figure \ref{fig_global_vod}(a) for $\beta=0.0$ and in Figure \ref{fig_global_vod}(b) for $\beta=0.5$. $V_{\phi}/\sigma_{R\star}$ represents the amount of ordered rotation in stellar kinematics and is abbreviated as $V/\sigma_\star$ from hereon. The average and standard deviation of $\Delta V_\mathrm{ADC}/V_\mathrm{CO}$ in each $V/\sigma_\star$ bin are listed in Table \ref{tab_vod}. In $0.5<V/\sigma_\star<3$, $V_\mathrm{ADC, \beta=0.0}$ underestimate $V_\mathrm{c}$ by up to to $\sim$18\% in a bin, with both an increasing $\Delta V_\mathrm{ADC}/V_\mathrm{CO}$ and an increasing scatter towards the low $V/\sigma_\star$ regime. $V_\mathrm{ADC, \beta=0.5}$ agrees better with $V_\mathrm{CO}$ in all the $V/\sigma_\star>1.0$ bins, with a difference averaging to $<4\%$ in this regime. For $V/\sigma_\star<1.0$, however, $V_\mathrm{ADC, \beta=0.5}$ overestimate $V_\mathrm{c}$ by 14$\%$. The scatter in $\Delta V_\mathrm{ADC}/V_\mathrm{CO}$ for the case of $\beta=0.5$ also increases towards the low $V/\sigma_\star$ regime. 

\begin{figure*}
\begin{center}
\includegraphics[trim=0 30 0 5, clip=true,width=0.9\textwidth,page=1,angle=90.]{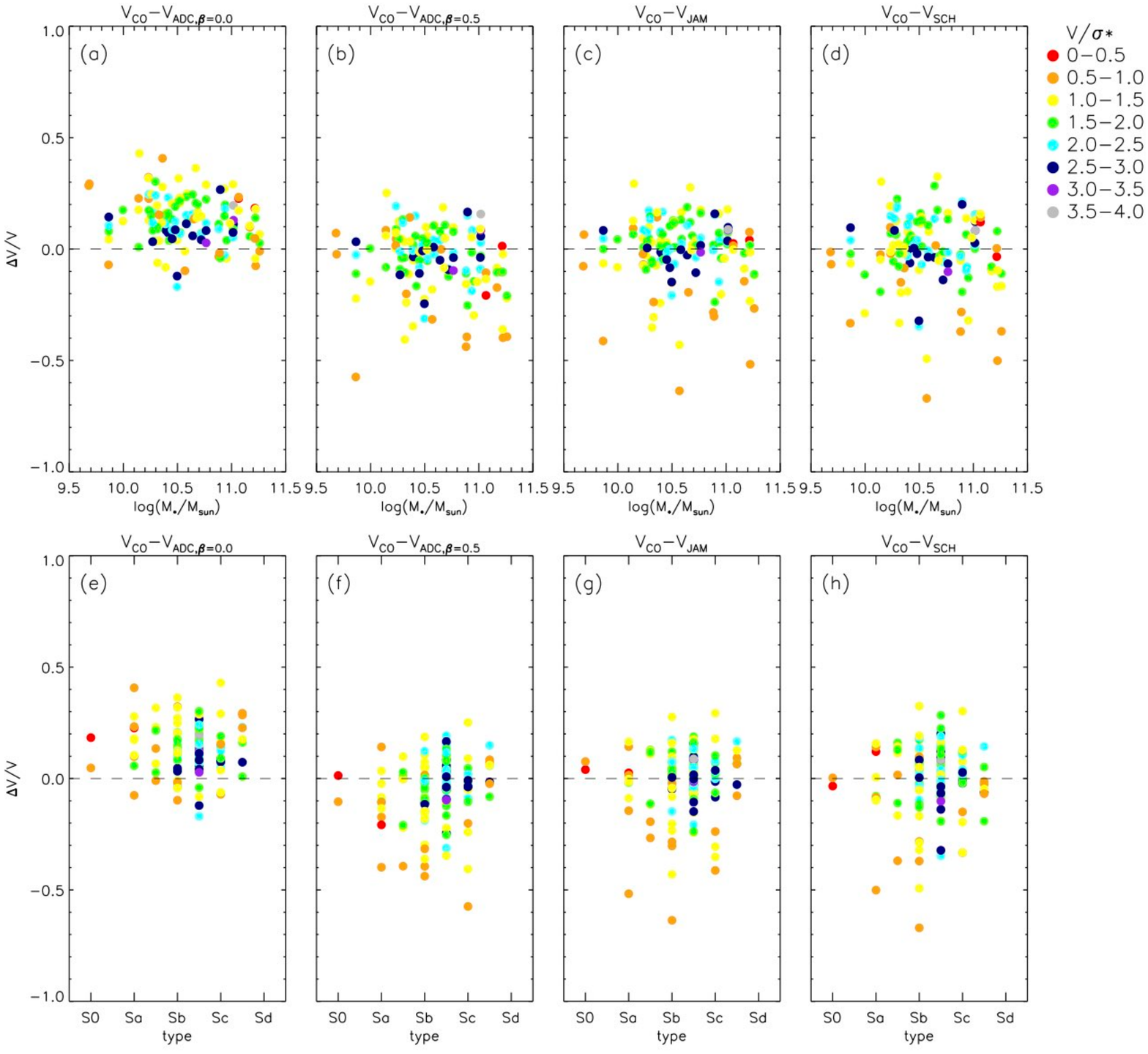}
\caption{Velocity differences between the stellar and gaseous circular velocity curves plotted against the total stellar mass (top row) and morphological types (bottom row) of the galaxies. Each dot here correspond to a grey dot in Figure \ref{fig_global_vod}(a), (b) for ADC, Figure \ref{fig_global_vod}(c) for JAM and Figure \ref{fig_global_vod}(d) for SCH. Only points with $R>3\sigma_\mathrm{beam}$ are included in these plots. The points are colour-coded with their respective $V/\sigma_\star$ value. No systematic trends can be found in $\Delta V/V$ with respect to galactic properties.}
\label{fig_global_par}
\end{center}
\end{figure*}

To discern any systematics in the difference between $V_\mathrm{ADC}$ and $V_\mathrm{CO}$ with galactic properties, we show plots of $\Delta V_\mathrm{ADC}/V_\mathrm{CO}$ against stellar mass and morphological types for $\beta=0.0$ in Figure \ref{fig_global_par}(a) and \ref{fig_global_par}(e), and for $\beta=0.5$ in Figure \ref{fig_global_par}(b) and \ref{fig_global_par}(f). Each circle correspond to one grey circle in Figure \ref{fig_global_vod}, colour coded here with the respective $V/\sigma_\star$ bin value, with the lowest $V/\sigma_\star$ bin ($0-0.5$) coloured red and the highest $V/\sigma_\star$ bin ($3.5-4.0$) coloured grey. We do not find any trends in $\Delta V_\mathrm{ADC}/V_\mathrm{CO}$ with respect to these galactic properties.

\subsection{JAM vs. CO}
The values of $V_\mathrm{JAM}$ are plotted against that of $V_\mathrm{CO}$ in Figure \ref{fig_reff}(c), and show good agreement with $V_\mathrm{CO}$ at $R=R_\mathrm{e}$. Again, we extrapolate $V_\mathrm{JAM}$ to 1\,$R_\mathrm{e}$ using the MGEs and show in open circles $V_\mathrm{c}$ at $R_\mathrm{max}$ for galaxies which have $R_\mathrm{max}<R_\mathrm{e}$. The corresponding histogram of $Q_\mathrm{JAM}=(1-\frac{V_\mathrm{JAM}}{V_\mathrm{CO}})_{R_\mathrm{e}}$ is shown in Figure \ref{fig_reff}(d). The mean and standard deviation of $Q_\mathrm{JAM}$ are $-$0.3\% and 8\% respectively, indicating a good agreement between $V_\mathrm{JAM}$ with $V_\mathrm{CO}$ at 1 $R_\mathrm{e}$, with no preferential bias (of either being smaller or larger than $V_\mathrm{CO}$). Already, this tells us that without the thin disk assumption, JAM can well recover $V_\mathrm{c}$. 

Again we show the relative difference $\Delta V_\mathrm{JAM}/V_\mathrm{CO}$ against $R/R_\mathrm{e}$ in Figure \ref{fig_global_r}(c). The average and standard deviation in each radial bin are listed in Table \ref{tab_r}. On average, $V_\mathrm{JAM}$ agrees with $V_\mathrm{CO}$ to within $1\sigma$ at all radii, the scatter in $\Delta V_\mathrm{JAM}/V_\mathrm{CO}$ increases towards the centre to up to 17\% for $R<0.4R_\mathrm{e}$. Plotting $\Delta V_\mathrm{JAM}/V_\mathrm{CO}$ against $V/\sigma_\star$ in Figure \ref{fig_global_vod}(c) shows similar features, $\Delta V_\mathrm{JAM}/V_\mathrm{CO}$ agrees to within 1$\sigma$ at all bins, with an increasing scatter towards the low $V/\sigma_\star$ regime. No specific trend is seen in $\Delta V_\mathrm{JAM}/V_\mathrm{CO}$ with respect to $V/\sigma_\star$. 

Despite $\Delta V_\mathrm{JAM}/V_\mathrm{CO}$ agrees to within 1$\sigma$ at all radial and $V/\sigma_\star$ bins, we see a large scatter in $\Delta V_\mathrm{JAM}/V_\mathrm{CO}$. In particular towards the inner and low $V/\sigma_\star$ region. 
Again, to better understand this scatter, we investigate how $\Delta V_\mathrm{JAM}/V_\mathrm{CO}$ changes with various galactic properties. In Figure \ref{fig_global_par}(c) and Figure \ref{fig_global_par}(g), we plot $\Delta V_\mathrm{JAM}/V_\mathrm{CO}$ against the total stellar mass and morphological type of each galaxy respectively. No systematic trend can be found with respect to these galactic properties.

\subsection{SCH vs. CO}
$V_\mathrm{SCH}$ show good agreement with $V_\mathrm{CO}$ at 1$R_\mathrm{e}$, as shown in the one-to-one plot of $V_\mathrm{SCH}$ against $V_\mathrm{CO}$ in Figure \ref{fig_reff}(e). The corresponding $Q_\mathrm{SCH}=(1-\frac{V_\mathrm{SCH}}{V_\mathrm{CO}})_{R_\mathrm{e}}$ is shown in Figure \ref{fig_reff}(f). $Q_\mathrm{SCH}$ has a mean and a standard deviation of $-$0.2\% and 9\% respectively, again showing no preferential bias towards being positive or negative.

We plot the relative difference $\Delta V_\mathrm{SCH}/V_\mathrm{CO}$ against $R/R_\mathrm{e}$ in Figure \ref{fig_global_r}(d), and then against $V/\sigma_\star$ in Figure \ref{fig_global_vod}(d). The average and standard deviation in each radial and $V/\sigma_\star$ bins are listed in Table \ref{tab_r} and Table \ref{tab_vod}. On average, $\Delta V_\mathrm{SCH}/V_\mathrm{CO}$ agrees to within $1\sigma$ at all radial bins. Just like JAM, the scatter in $\Delta V_\mathrm{SCH}/V_\mathrm{CO}$ also increases towards the centre up to 17\% for $R<R_\mathrm{e}$. Also, no systemic trend is seen with respect to $V/\sigma_\star$ values. We investigate how $\Delta V_\mathrm{SCH}/V_\mathrm{CO}$ varies with respect to total stellar mass in Figure \ref{fig_global_par}(d) and morphological type in Figure \ref{fig_global_par}(h) but once again find no systematic trend.

\subsection{Comparison between the three stellar dynamical models}
\begin{figure}
\includegraphics[trim=10 30 80 30, clip=true,width=0.48\textwidth]{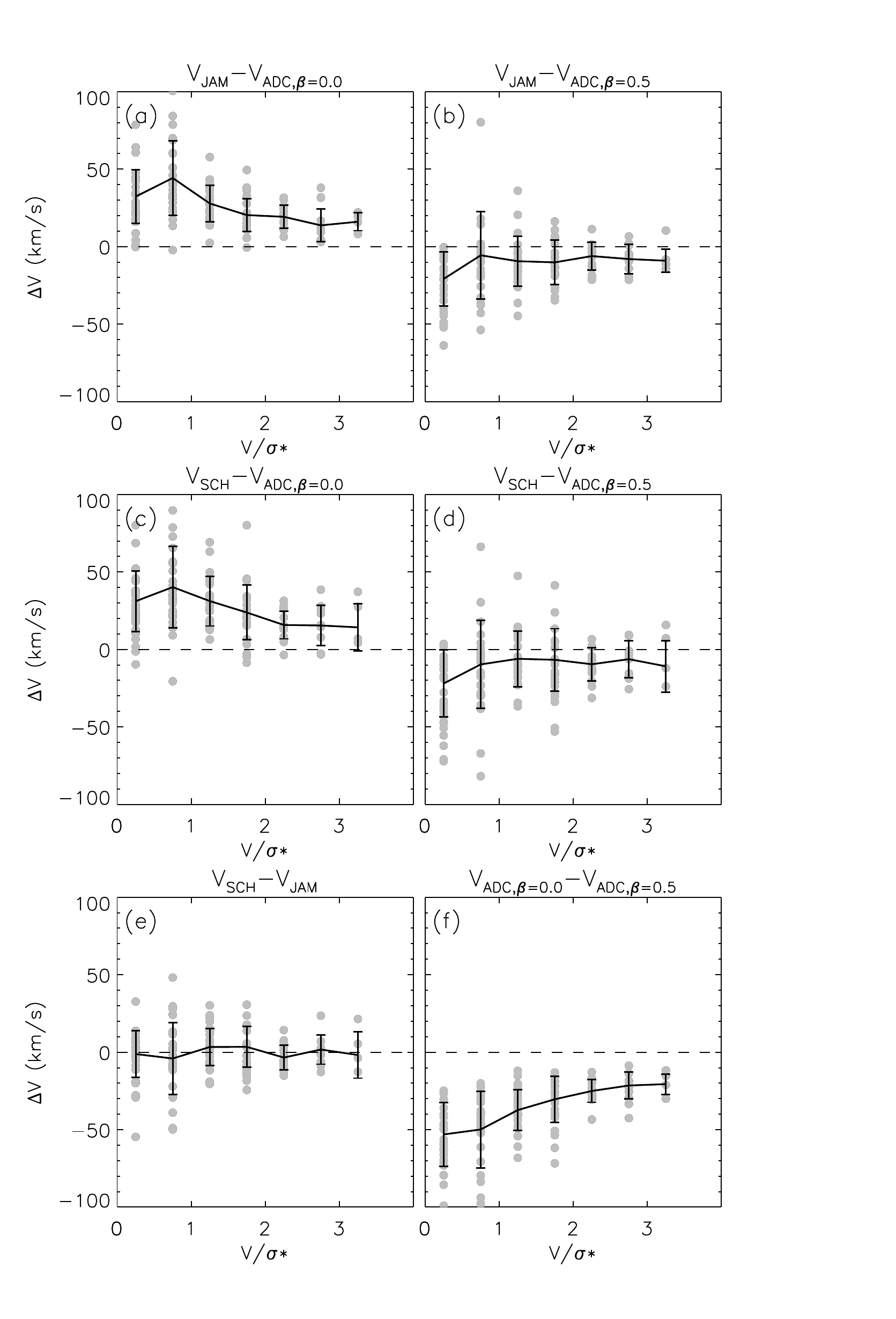}
\caption{Comparison between $V_\mathrm{c}$ extracted from stellar kinematics using JAM, ADC and SCH at different $V/\sigma_{\star}$. Each grey dots represent a measurement from one galaxy at that specific $V/\sigma_\star$ bin. The black curve show the mean and the error bars show the standard deviation of each bin. ADC shows a smaller $V_\mathrm{c}$ when compared to either JAM or SCH, and the differences increases towards lower $V/\sigma_\star$. Comparing JAM and SCH also shows a slight trend: in the low $V/\sigma_\star$ regime, JAM tends to produce $V_\mathrm{c}$ that are higher than SCH while in the high $V/\sigma_\star$ regime, JAM tends to produce $V_\mathrm{c}$ that are lower than SCH.}
\label{fig_diff_star}
\end{figure}

In Figure \ref{fig_diff_star}, we show the differences between the circular velocity obtained using the three different methods using the same stellar kinematics. Each grey dot correspond to the velocity difference measured at a certain $V/\sigma_\star$ bin of a galaxy. With the black curve and corresponding error bars we show the average and standard deviations of the differences in stellar $V/\sigma_\star$ bins, we list the corresponding values in Table \ref{tab_diff_star}. 

Comparing the two models that derive $V_\mathrm{c}$ by solving the Jeans equation, ADC and JAM (Figure \ref{fig_diff_star}(a) and \ref{fig_diff_star}(b)), shows that $V_\mathrm{ADC, \beta=0.0}$ in general are smaller than $V_\mathrm{JAM}$. Moreover, the difference between the two increases with decreasing $V/\sigma_\star$, the same trend had been found with SAURON late-type spiral galaxies in \cite{kal17}. Especially at the regime $V/\sigma_\star<1$, where the random motion dominate over the ordered rotation, the difference between ADC and JAM reaches an average of $\sim$36\,km\,s$^{-1}$. $V_\mathrm{ADC, \beta=0.5}$, on the other hand, agrees with $V_\mathrm{JAM}$ to within 1\,$\sigma$ at all $V/\sigma_\star$ bins $>0.5$. In the lowest $V/\sigma_\star$ bin of $V/\sigma_\star<0.5$, however, $V_\mathrm{ADC, \beta=0.5}$ is larger than $V_\mathrm{JAM}$ on average by $\sim$21\,km\,s$^{-1}$. 

We next compare $V_\mathrm{SCH}$ and $V_\mathrm{ADC}$ in Figure \ref{fig_diff_star}(c) and \ref{fig_diff_star}(d). Just like when compared with $V_\mathrm{JAM}$, $V_\mathrm{ADC, \beta=0.0}$ is smaller than $V_\mathrm{SCH}$, with an increasing difference towards lower $V/\sigma_\star$ to on average by $\sim$33\,km\,s$^{-1}$ at $V/\sigma_\star<1$. 
$V_\mathrm{ADC, \beta=0.5}$, on the other hand, agrees with $V_\mathrm{SCH}$ to within 1\,$\sigma$ on average except for the $V/\sigma_\star<0.5$ bin. There, $V_\mathrm{ADC, \beta=0.5}$ is larger than $V_\mathrm{SCH}$ by $\sim$22\,km\,s$^{-1}$ on average.

Both the Jeans and Schwarzschild methods take into account the full line-of-sight integration when modelling the observed mean velocity and velocity dispersion map. The two models show good agreement to within 4$\%$ bins on average, with scatters of $\sim$8-23$\%$. 

The biggest difference is shown when comparing the two $V_\mathrm{c}$ derived from ADC, with $\beta=0.0$ and $\beta=0.5$, as shown in Figure \ref{fig_diff_star}(f). $V_\mathrm{ADC, \beta=0.0}$ is always smaller than $V_\mathrm{ADC, \beta=0.5}$, with the average difference increasing towards lower $V/\sigma_\star$ regimes up to $>$50\,km\,s$^{-1}$.

\begin{table*}
\begin{tabular}{@{}lllllllll}
\hline
$R/R_\mathrm{e}$ & $\overline{\Delta V_\mathrm{ADC}/V}$ & $\sigma_\mathrm{\Delta V/V,ADC}$ & $\overline{\Delta V_\mathrm{ADC}/V}$ & $\sigma_\mathrm{\Delta V/V,ADC}$ & $\overline{\Delta V_\mathrm{JAM}/V}$ & $\sigma_\mathrm{\Delta V/V,JAM}$ & $\overline{\Delta V_\mathrm{SCH}/V}$ & $\sigma_\mathrm{\Delta V/v,SCH}$ \\
& $\beta=0.0$ & $\beta=0.0$ & $\beta=0.5$ & $\beta=0.5$ &&&& \\
\hline
\hline

0.2-0.4 & 0.19 & 0.12 & $-$0.03 & 0.16 &      0.02 & 0.17 & 0.00 & 0.17\\
0.4-0.6 & 0.16 & 0.08 & $-$0.02 & 0.11 &      0.03 & 0.11 & 0.01 & 0.12\\
0.6-0.8 & 0.13 & 0.07 & $-$0.02 & 0.10 &      0.01 & 0.08 & 0.00 & 0.09\\
0.8-1.0 & 0.12 & 0.06 & $-$0.02 & 0.09 &      0.01 & 0.06 & 0.00 & 0.08\\
1.0-1.2 & 0.07 & 0.05 & $-$0.07 & 0.07 & $-$0.02 & 0.05 & $-$0.05 & 0.06\\
1.2-1.4 & 0.08 & 0.05 & $-$0.07 & 0.05 & $-$0.04 & 0.07 & $-$0.04 & 0.12\\
1.4-1.6 & 0.02 & 0.05 & $-$0.14 & 0.07 & $-$0.06 & 0.04 & $-$0.05 & 0.11\\
\hline
\end{tabular}
\caption{Discrepancies in derived circular velocities between CO and stellar kinematics, listed in bins of $R/R_\mathrm{e}$. $\overline{\Delta V}$ is the error-weighted average of stellar and gaseous velocity difference in each bin, $\sigma_\mathrm{\Delta V}$ is the corresponding standard deviation. All velocities are listed in the unit of $\mathrm{km\,s^{-1}}$. \label{tab_r}}
\vspace{2mm}

\begin{tabular}{@{}lllllllll}
\hline
$V/\sigma_\star$ & $\overline{\Delta V_\mathrm{ADC}/V}$ & $\sigma_\mathrm{\Delta V/V,ADC}$ & $\overline{\Delta V_\mathrm{ADC}/V}$ & $\sigma_\mathrm{\Delta V/V,ADC}$ & $\overline{\Delta V_\mathrm{JAM}/V}$ & $\sigma_\mathrm{\Delta V/V,JAM}$ & $\overline{\Delta V_\mathrm{SCH}/V}$ & $\sigma_\mathrm{\Delta V/V,SCH}$\\
& $\beta=0.0$ & $\beta=0.0$ & $\beta=0.5$ & $\beta=0.5$ &&&& \\
\hline
\hline

0.5-1.0 & 0.14 & 0.12 & $-$0.14 & 0.15 & $-$0.10 & 0.18 & $-$0.04 & 0.21 \\
1.0-1.5 & 0.18 & 0.09 & $-$0.01 & 0.11 &      0.03 & 0.11 &       0.00 & 0.12 \\
1.5-2.0 & 0.11 & 0.06 & $-$0.04 & 0.09 &      0.01 & 0.07 & $-$0.01 & 0.09 \\
2.0-2.5 & 0.12 & 0.07 &      0.00 & 0.09 &      0.03 & 0.06 & $-$0.01 & 0.08 \\
2.5-3.0 & 0.08 & 0.04 & $-$0.04 & 0.05 &      0.00 & 0.07 & $-$0.01 & 0.09 \\
3.0-3.5 & 0.19 & 0.07 &      0.03 & 0.12 &      0.06 & 0.06 &      0.03 & 0.12 \\
\hline
\end{tabular}
\caption{Discrepancies in derived circular velocities between CO and stellar kinematics, listed in bins of stellar velocity dispersion, $V/\sigma_\star$. $\overline{\Delta V}$ is the error-weighted average of stellar and gaseous velocity difference in each bin, $\sigma_\mathrm{\Delta V}$ is the corresponding standard deviation. All velocities are listed in the unit of $\mathrm{km\,s^{-1}}$. \label{tab_vod}}
\vspace{2mm}

\begin{tabular}{@{}lllllllllllll}
\hline
 & \multicolumn{2}{c} {$V_\mathrm{SCH-ADC,\beta=0.0}$} & \multicolumn{2}{c} {$V_\mathrm{SCH-ADC,\beta=0.5}$}  & \multicolumn{2}{c} {$V_\mathrm{JAM-ADC, \beta=0.0}$ } & \multicolumn{2}{c} {$V_\mathrm{JAM-ADC, \beta=0.5}$} & \multicolumn{2}{c}{ $V_\mathrm{SCH-JAM}$ } & \multicolumn{2}{c}{$V_\mathrm{ADC, \beta=0.0-\beta=0.5}$}\\
$V/\sigma_{\star}$ & $\overline{\Delta V}$ & $\sigma_{\Delta V}$ & $\overline{\Delta V}$ & $\sigma_{\Delta V}$ & $\overline{\Delta V}$ & $\sigma_{\Delta V}$ & $\overline{\Delta V}$ & $\sigma_{\Delta V}$ & $\overline{\Delta V}$ & $\sigma_{\Delta V}$ & $\overline{\Delta V}$ & $\sigma_{\Delta V}$\\
\hline
\hline

0.0-0.5 & 31.1 & 19.4 & $-$22.0 & 21.6 & 32.3 & 17.2 & $-$20.8 & 17.5 & $-$1.18 & 15.1 & $-$53.1 & 20.6 \\
0.5-1.0 & 40.2 & 26.2 & $-$9.67 & 28.3 & 44.3 & 24.2 & $-$5.59 & 28.2 & $-$4.08 & 23.1 & $-$49.9 & 24.7 \\
1.0-1.5 & 31.2 & 16.1 & $-$6.08 & 18.0 & 27.9 & 11.7 & $-$9.43 & 16.2 & $  $3.35 & 11.9 & $-$37.4 & 13.2 \\
1.5-2.0 & 23.8 & 17.6 & $-$6.71 & 20.1 & 20.4 & 10.5 & $-$10.2 & 14.4 & $  $3.46 & 13.3 & $-$30.4 & 14.8 \\
2.0-2.5 & 15.8 & 8.96 & $-$9.59 & 10.7 & 19.2 & 7.44 & $-$6.11 & 8.97 & $-$3.47 & 7.97 & $-$25.0 & 7.31 \\
2.5-3.0 & 15.4 & 13.1 & $-$6.31 & 11.9 & 13.7 & 10.5 & $-$8.03 & 9.46 & $  $1.72 & 9.49 & $-$21.5 & 8.57 \\
3.5-4.0 & 14.3 & 15.1 & $-$10.9 & 16.4 & 16.1 & 5.76 & $-$9.11 & 7.53 & $-$1.81 & 15.0 & $-$20.6 & 6.56 \\
\hline
\end{tabular}
\caption{Discrepancies in derived circular velocities with different models using stellar kinematics, listed in bins of stellar velocity dispersion, $V/\sigma_\star$. $\overline{\Delta V_\mathrm{JAM-ADC}}$, $\overline{\Delta V_\mathrm{SCH-JAM}}$ and $\overline{\Delta V_\mathrm{SCH-ADC}}$ are the error-weighted average of stellar and gaseous velocity difference in each bin, $\sigma_\mathrm{JAM-ADC}$, $\sigma_\mathrm{SCH-JAM}$ and $\sigma_\mathrm{SCH-ADC}$ are the corresponding standard deviation. All velocities are listed in the unit of $\mathrm{km\,s^{-1}}$. \label{tab_diff_star}}
\end{table*}

\section{Discussion}\label{sect_diss}

In this section, we discuss the possible reasons for the disagreements we see between the $V_\mathrm{c}$ obtained from different stellar dynamical models and CO, as well as their respective trends with radius and galactic properties. To recap, we find that: (1) $V_\mathrm{ADC, \beta=0.0}$ underestimate $V_\mathrm{c}$ by $\sim$8-20\%, showing a trend of increasing relative difference $\Delta V/V$ with respect to the $V_\mathrm{CO}$, as well as scatter in $\Delta V/V$, towards the inner region. (2) On average, $V_\mathrm{ADC, \beta=0.5}$, $V_\mathrm{JAM}$ and $V_\mathrm{SCH}$ agree with CO to within $1\sigma$ over all radii. 
(3) Towards the inner region ($R<0.4R_\mathrm{e}$) and low $V/\sigma_\star(<1)$ regime, we find a large scatter among our galaxy sample of 15\%, 18\% and 21\% in $\Delta V/V$, for $V_\mathrm{ADC, \beta=0.5}$, $V_\mathrm{JAM}$ and $V_\mathrm{SCH}$ respectively. (4) Within the large scatter, we do not find any systematic trend with respect to galactic properties such as stellar mass and morphological type. All of these comparisons are done with data outside of 3$\sigma_\mathrm{beam}$ of the CO observations. One should keep in mind that part of the scatter arises from the noise in both the CO and the stellar kinematics ($\sim$5\% in the innermost region).
Comparing the $V_\mathrm{c}$ obtained from the 3 stellar dynamical models directly with each other gives three main results: (1) $V_\mathrm{ADC, \beta=0.0}$ is smaller than $V_\mathrm{ADC, \beta=0.5}$, $V_\mathrm{JAM}$ and $V_\mathrm{SCH}$, with differences increasing towards lower $V/\sigma_\star$, 
(2) while $V_\mathrm{ADC,\beta=0.5}$ agree with both $V_\mathrm{JAM}$ and $V_\mathrm{SCH}$ at $V/\sigma_\star>0.5$ to within 1\,$\sigma$, it is on average larger than both by $\sim$20\,km\,s$^{-1}$ at $V/\sigma_\star<0.5$, and (3) that $V_\mathrm{JAM}$ and $V_\mathrm{SCH}$ are in excellent agreement with each other. 

\subsection{Effects of model assumptions on derived $V_\mathrm{c}$}

The ADC models assume a thin disk distribution of stars and therefore cannot account for masses distributed away from the $z=0$ plane. This is the case for  $V_\mathrm{ADC, \beta=0.0}$, which underestimate $V_\mathrm{c}$ at all radii. The trend of velocity discrepancies with radius can also be explained by the fact that thick disks and/or bulges in galaxies tend to be more prominent in the inner region, both of which imply masses distributed away from the disk plane and hence reduces the accuracy of the ADC model. We show however, that by adopting $\beta=0.5$, the ADC models can reproduce accurate $V_\mathrm{c}$. 
Such agreement is not surprising when one consider that the light-weight kinematic measurements are mostly dominated by young bright stars which lie close to the disk plane. 
We should emphasis here that the agreement between $V_\mathrm{ADC, \beta=0.5}$ with $V_\mathrm{CO}$ does not suggest that the intrinsic value of $\beta$ is 0.5, but rather, under the incomplete (thin disk) assumption of ADC, $\beta=0.5$ can empirically provide a good estimate of the true $V_\mathrm{c}$ except for the high mass galaxies (with $V_\mathrm{c}\gtrsim280$\,km\,s$^{-1}$). Similar overestimation of $V_\mathrm{c}$ can be seen in at low $V/\sigma_\star$ ($<$1) by $V_\mathrm{ADC, \beta=0.5}$. This might indicate that in rounder and hotter systems such as early type high-mass galaxies or the inner region of disk galaxies, assuming  $\beta=0.5$ is an overkill even when adopting the thin-disk assumption, as such systems are likely to be more isotropic. The similar differences in the derived $V_\mathrm{c}$ at $V/\sigma_\star<0.5$ seen when $V_\mathrm{ADC, \beta=0.5}$ is compared with $V_\mathrm{JAM}$ and $V_\mathrm{SCH}$ are likely caused by the same reason.


Since both the Jeans and Schwarzschild models take into account the full line-of-sight integration of the stellar kinematics, masses distributed away from the disk plane can also be taken into account in these models. The good agreement between $V_\mathrm{JAM}$ and $V_\mathrm{SCH}$ with $V_\mathrm{CO}$ at $R>0.5R_\mathrm{e}$ suggests that both models are reliable in recovering the dynamical masses of galaxies at larger radii. For the inner region, the large scatter between $V_\mathrm{JAM}$ or $V_\mathrm{SCH}$ and $V_\mathrm{CO}$ suggests, however, that one should be aware of the possible discrepancies when interpreting the result from the models in these regimes. 

Below we suggest the possible reasons causing the $\sim$20$\%$ scatter in both the Jeans and Schwarzschild models when being compared with CO in the innermost region. The stellar mass-to-light ratio is still assumed to be constant in both models. Stellar mass-to-light ratio tends to increase toward the inner region due to the increasing stellar age. How the two models compensate for the incorrectly estimated stellar mass with the dark matter component would affect the resultant total mass-to-light ratio. In addition, the assumed shape of the underlying mass distribution can also affect the resulting $V_\mathrm{c}$. In particular, we assume a spherical dark matter halo and that the stellar mass distribution follows the shape of light distribution. If the mass distribution assumed is flatter than the true distribution, one would overestimate the $V_\mathrm{c}$ and vice versa \citep{bin87}. In galaxies, the older stars that are scattered higher above the disk plane would have a higher M/L ratio than the younger stars in the disk plane, leading to a less flattened distribution in mass compared to light. Although both effects should be more prominent in the inner region of the older galaxies and rounder systems such as the earlier type galaxies, the opposite effects can wash out any systematic trend in the discrepancies with galaxy types.

We would like to warn our readers that even though JAM reproduces $V_\mathrm{c}$ in good agreement with CO (at $R\gtrsim R_\mathrm{e}$ and high $V/\sigma_\star$ regimes) or the Schwarzschild models, the other extracted parameters such as $\beta_\mathrm{z}$ or mass ratio between dark matter and luminous matter are not necessarily correct or physical. This has been reflected by the few galaxies with which $\beta_\mathrm{z}$ and $V_\mathrm{vir}$ reach the boundaries of the parameter space to unphysical values. In both cases, Schwarzschild models provide well constrained $\beta_\mathrm{z}$ and $V_\mathrm{vir}$. The inability of JAM in recovering $\beta_\mathrm{z}$ and $V_\mathrm{vir}$ in certain galaxies is likely caused by the fact that these galaxies do not satisfy additional assumptions in JAM models, such as having velocity ellipsoids aligned with the cylindrical coordinate system. 

\subsection{Implications on high redshift Tully-Fisher relation}
The evolution of the Tully-Fisher relation towards high redshift, $z$, is a subject of debate. While some authors find no significant evolutions \citep[e.g.][]{mill11,mon17,pell17}, others find an evolution towards a lower zero-point (in stellar mass) at high-$z$ \citep[e.g.][]{cre09,til16,pri16}. When obtaining the rotation velocity from high-redshift galaxies, emission lines from ionised gas are often used as the kinematic tracer. Such high-$z$ ionised gas kinematics show similar $V/\sigma$ values as the local stellar kinematics in our sample \citep[$\sim0.5-4$ at $z\sim2$,][]{wis15}. Various authors took different approaches in dealing with the high dispersion of the ionised gas kinematics at high-$z$, namely, either by disregarding the galaxy with low $V/\sigma$ in their sample, or by taking an approximated form of $V_\mathrm{c}$ such as $V_\mathrm{rms}=\sqrt{V^{2}+\sigma^{2}}$. Our results suggest that one of the three models can be taken to recover $V_\mathrm{c}$ from the high-dispersion ionised gas kinematics at high-$z$.

\section{Summary}\label{sect_sum}
Stars are present in all galaxies and can serve as a kinematic tracer for the underlying dynamical masses. The collisionless nature of stellar orbits, however, renders such task non-trivial and various dynamical models have been developed to solve the problem. In this paper, we test the validity of three commonly used stellar dynamical models in recovering the underlying total mass in galaxies by comparing the circular velocities ($V_\mathrm{c}$) obtained from IFU stellar kinematics to that extracted from cold molecular gas kinematics over a large and homogeneous sample of 54 galaxies. Such comparison is for the first time enabled by two large surveys of nearby galaxies: the EDGE and the CALIFA survey. We extracted cold gas rotation curves from the CARMA EDGE survey CO$(J=1-0)$ line emission. We applied harmonic decomposition to the mean velocity fields to remove perturbations from, for example, a bar or spiral arms. For the same galaxies, we show $V_\mathrm{c}$ obtained from stellar kinematics from the CALIFA survey, using the Asymmetric Drift Correction (ADC), Axisymmetric Jeans Anisotropic Multi-gaussian expansion Models (JAM) and Schwarzschild (SCH) models. For ADC, we tested the model with two commonly adopted constant velocity anisotropy values: $\beta=0.0$ (isotropic) and $\beta=0.5$. For JAM, we assume a constant anisotropy, a constant stellar mass-to-light ratio and a spherical NFW dark matter halo, which are obtained from fitting the velocity moments. The Schwarzschild models adopt an orbit-based approach, with which we again model both the luminous (assuming a constant mass-to-light ratio) and dark matter masses (with an NFW halo), but with no assumption on the velocity anisotropy. 

We compare the circular velocities obtained from kinematically cold molecular gas CO with that obtained from stellar kinematics. At the effective radii ($R_\mathrm{e}$), all the anisotropic ADC ($\beta=0.5$), JAM and Schwarzschild models reproduce $V_\mathrm{CO}$ to within $<$5\%, with scatter $<$10\%. Specifically, $Q_\mathrm{ADC, \beta=0.5}=-5\pm11\%$, $Q_{JAM}=-0.3\pm11\%$ and $Q_\mathrm{SCH}=-0.2\pm14\%$ (where $Q_\mathrm{X}=1-\frac{V_\mathrm{x}}{V_\mathrm{CO}}$). In the inner regions ($R<0.4R_\mathrm{e}$), the scatter increases to $\sim20\%$ for all methods.

The excellent performance of even ADC, which has the strictest assumptions, is likely due to the luminosity weighted velocities in our IFU data - for which the brightest youngest stellar component will be predominantly the dynamically coldest and thinnest.

Possible reasons leading to such discrepancies between the stellar and CO $V_\mathrm{c}$ in the inner regions are as follows. ADC assumes stars to lie on a thin disk on the plane $z=0$, therefore it cannot capture masses distributed away from this plane. In particular, in the inner region, the presence of a bulge or a thick disk would render the ADC models to underestimate the circular velocities even more, as reflected by the increasing discrepancies between the $V_\mathrm{ADC, \beta=0.0}$ and $V_\mathrm{CO}$ towards the inner region. By assuming $\beta=0.5$, ADC can empirically recover $V_\mathrm{c}$ for galaxies with $V_\mathrm{c}<280$\,km\,s$^{-1}$.

Both the JAM and Schwarzschild models account for the 3 dimensional distribution of mass, 
however we suggest that the reasons for $\sim$20\% scatter in the relative difference between both models and $V_\mathrm{CO}$ in the inner region to be: (1) the deviation of the fitted constant stellar mass-to-light ratio to the intrinsic radially varying value, and (2) the possibility that the underlying shape of the dark matter and stellar mass distribution differ from the assumed shape of spherical halo and light distribution respectively.

This work shows therefore that accurate dynamical masses for galaxies can be recovered from modelling the integrated stellar kinematics with these three methods. Since $V_\mathrm{ADC, \beta=0.0}$ underestimate $V_\mathrm{c}$ by $\sim$12-20$\%$ at $R<R_\mathrm{e}$, we advise that this method is least suitable -instead, the ADC method can still be applied using $\beta=0.5$ which give a compatible estimate for $V_\mathrm{c}$ to within $\sim$10\% at $R_\mathrm{e}$. 
Significant deviations in the recovered values still possible locally due to non-constant baryonic and dark mass distributions, we hence advise readers to be aware of such possible discrepancies when interpreting the results from stellar dynamical models.

\section*{Acknowledgments}
This study uses data provided by the Calar Alto Legacy Integral Field Area (CALIFA) survey (http://califa.caha.es/) and the CARMA Extragalactic Database for Galaxy Evolution (EDGE) survey (http://www.astro.umd.edu/EDGE/). We would like to thank the EDGE collaboration for useful discussions which helped improve this manuscript.  This work was supported by Sonderforschungsbereich SFB 881 "The Milky Way System" (subproject A7 and A8) of the Deutsche Forschungsgemeinschaft (DFG). RL was supported by funding from the Natural Sciences and Engineering Research Council of Canada PDF award. RL, GvdV and JF-B. acknowledge support from grant AYA2016-77237-C3-1-P from the Spanish Ministry of Economy and Competitiveness (MINECO). ADB and RCL wish to acknowledge partial support from grants NSF-AST1412419 and NSF-AST1615960. LB and DU are supported by the National Science Foundation (NSF) under grants AST-1616924. This work was supported by the DAGAL network from the People Programme (Marie Curie Actions) of the European Union's Seventh Framework Programme FP7/2007- 2013/ under REA grant agreement number PITN-GA-2011-289313.

\bibliographystyle{mn2e}
\bibliography{examplerefs}   

\appendix

\section{Beam-smearing correction on CO mean velocity and velocity dispersion fields}\label{app_beamcorr}
\begin{figure}
\begin{center}
\includegraphics[trim=220 120 200 120, clip=true,width=0.45\textwidth]{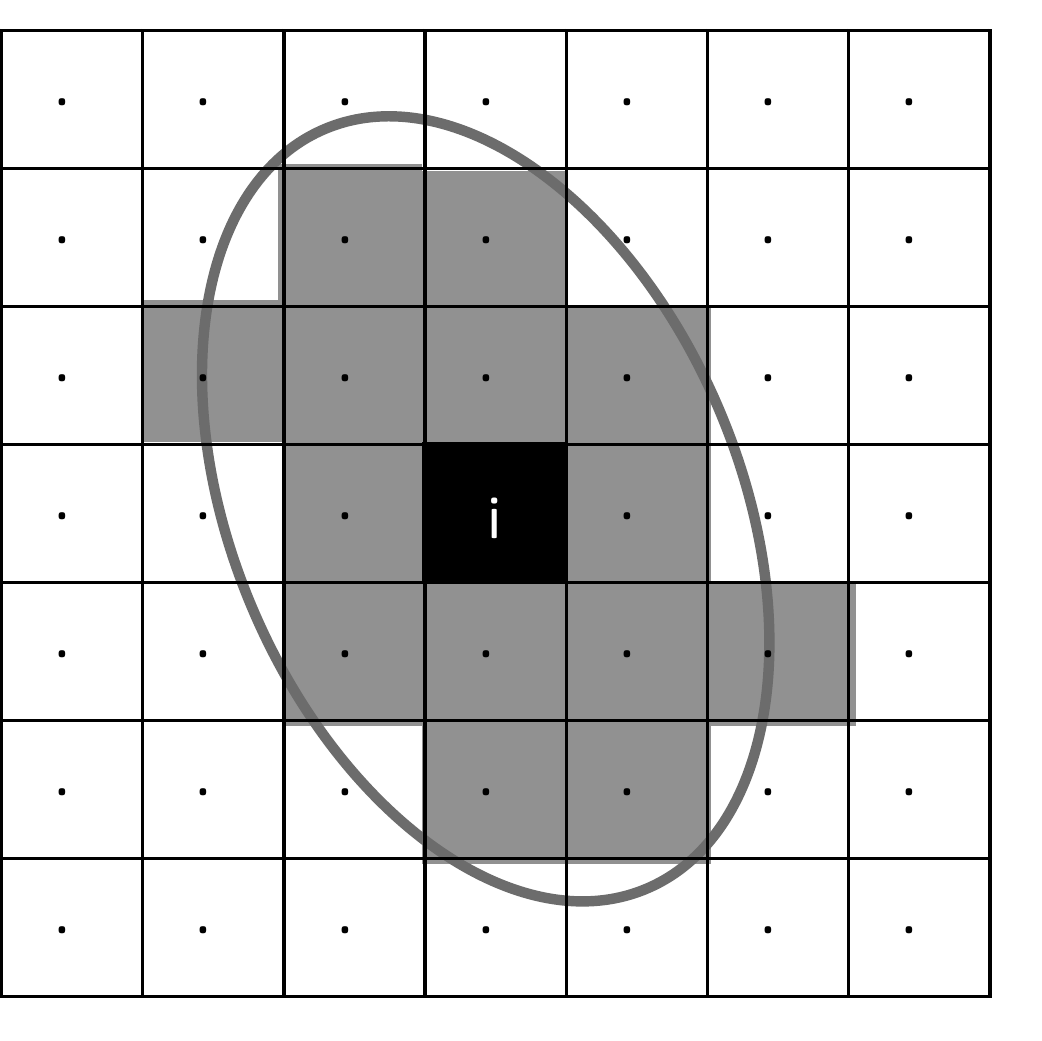}
\caption{Illustration of selected pixel set $X_{i}$ (in grey) around pixel $i$ (in black). The grey ellipse denotes two FWHM of the beam.}
\label{fig_bsc}
\end{center}
\end{figure}

The observed mean velocity field (especially in the inner region) as well as the velocity dispersion field are affected by beam smearing effect as the observations have an average beam size of $\sim$4.5$\arcsec$. To recover the intrinsic $V_\mathrm{\phi}$ and $\sigma_\mathrm{CO}$, we need to estimate and remove the effect of beam smearing on dispersion. This is done in two steps: (1) recover the pre-beam-smeared mean velocity map and (2) calculate the velocity dispersion caused by the beam around each pixel from the pre-beam-smeared velocity field.

To recover the pre-beam-smeared mean velocity field, we assume the galaxy is a thin disk such that the mean velocity equals the line-of-sight velocity. We first create a perturbed velocity field $V'$ by varying the velocity at each pixel, within the range $V_\mathrm{obs} \pm \sigma_\mathrm{obs}$ (observed velocity dispersion). From $V'$ we calculate a modelled velocity field $V_\mathrm{mod}$ from the beam weighted average of the velocities within the two FWHM of the beam around each pixel. This is illustrated in Figure \ref{fig_bsc}, where the black pixel labelled as pixel i is the pixel at which we want to evaluate the beam-smearing corrected mean velocity, grey ellipse indicates two FWHM of the beam and the grey pixels indicate the pixels with which we compute the beam weighed average. 

We iterate on this procedure until a $V'$ field is found such that its model velocity $V_\mathrm{mod}$, reproduces the original, beam-smeared observed velocity field, $V_\mathrm{obs}$. This $V'$ is then taken as the intrinsic beam-smearing corrected mean velocity field, $V_\mathrm{int}$, and is related to the observed velocity field as:
\begin{equation}
V_\mathrm{obs,i} = \frac{\sum_{j \in X_{i}} w_\mathrm{ij} V_\mathrm{int,j}} {\sum_{j \in X_{i}} w_\mathrm{ij} },
\end{equation}
with $X_\mathrm{i}$ is the set of pixels within a full beam around pixel i (i.e. the grey pixels in Figure \ref{fig_bsc} and $w_\mathrm{ij}$ being the weight of the beam of pixel i (a 2D gaussian) at the $j$th pixel. This relation holds simultaneously for all pixels.

From $V_\mathrm{int}$ we can then compute $\sigma_\mathrm{mod}$, the amount of dispersion contributed from beam-smearing. First we take $\sigma_\mathrm{mod}$ at a certain pixel as the beam-weighted standard deviation of $V_\mathrm{int}$ within a full beam size around the pixel:
\begin{equation}
\sigma_\mathrm{mod,i} = \sqrt{\frac{\sum_{j \in X_{i}} w_\mathrm{ij} (V_\mathrm{int,j}-\overline{V_\mathrm{int,j}})^{2}} {\sum_{j \in X_{i}} w_\mathrm{ij} }}.
\end{equation}
 
Finally, we obtain the intrinsic dispersion, $\sigma_\mathrm{int}$ by performing a quadrature subtraction of the modelled dispersion, $\sigma_\mathrm{mod}$, from the observed dispersion, $\sigma_\mathrm{obs}$:
\begin{equation}
\sigma_\mathrm{int} = \sqrt{\sigma_\mathrm{obs}^{2}-\sigma_\mathrm{mod}^{2}}.
\end{equation}

\begin{figure}
\includegraphics[trim=280 320 0 0, clip=true,width=0.45\textwidth,angle=90]{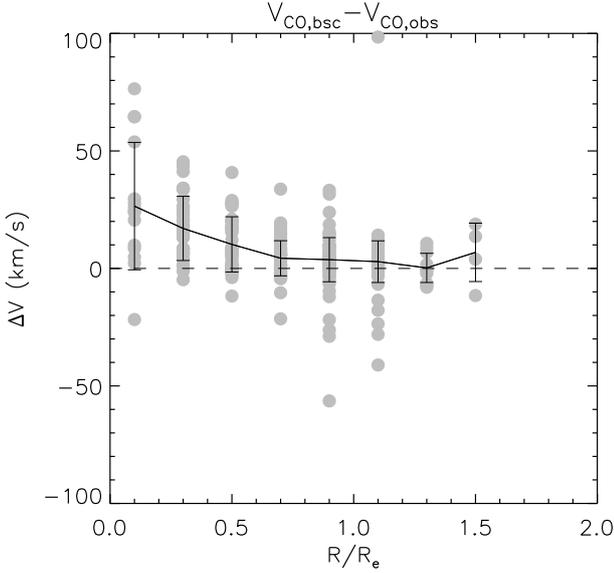}
\caption{We show the differences between the CO rotation curves before and after beam-semaring correction of all the 54 galaxies in grey dots. The black curve and error bars indicate the error-weighted mean and standard deviation of the differences in each bin. As shown in the plot, beam-smearing correction increases the rotation curve and the differences between the two increases inwards.}
\label{fig_bscvmod}
\end{figure}

\begin{figure*}
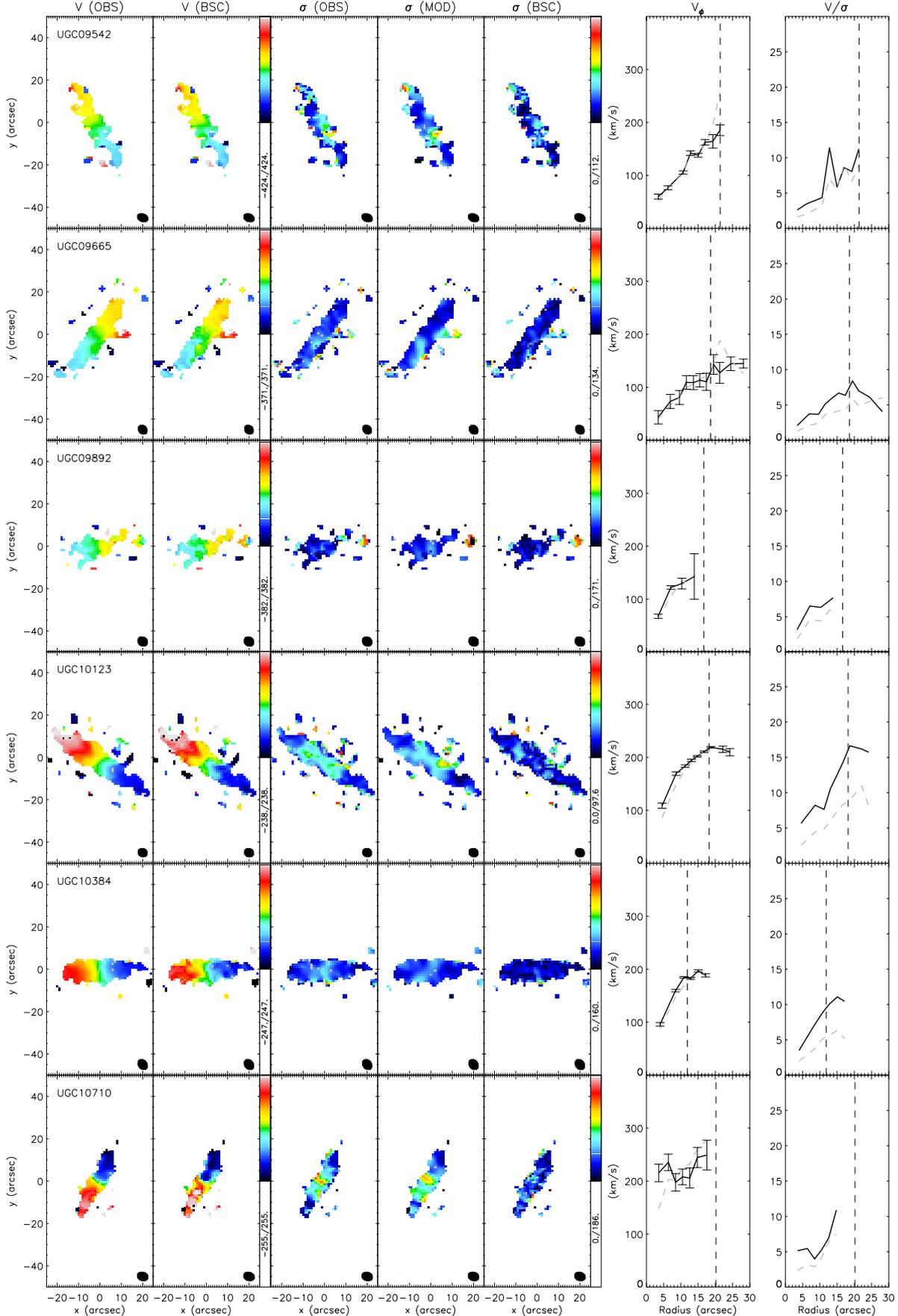

\centering 
\includegraphics[trim=73 40 55 20, page=17,width=0.6\textwidth,angle=90]{plotvod_all.pdf}
\includegraphics[trim=100 40 28 20, page=18,width=0.6\textwidth,angle=90]{plotvod_all.pdf}
\vspace*{15mm}
\caption{The seven figures for each galaxy from left to right are: (1) observed velocity field, (2), beam-smearing corrected velocity field, (3) observed dispersion field, (4) modelled dispersion field, (5) beam-smeared corrected dispersion field, all colour-coded in units of km\,s$^{-1}$; (6) CO rotation curve and (7) $V/\mathrm{\sigma}_\mathrm{CO}$ plot, where the grey line represent the observed value and the solid black line represent the corrected value, and the vertical dashed line marks the effective radius.}
\label{fig_plotvod_all}
\end{figure*}

In Figure\ref{fig_bscvmod}, we show the differences between the rotation curves extracted from the CO kinematics before and after beam-smearing correction. After beam-smearing correction, the rotation curve show a larger value, the differences may be negligible in the outer radii but become significant in the inner region where the gradient in the velocity field is larger. In Figure \ref{fig_plotvod_all} we show the observed velocity and dispersion fields of all the 54 galaxies in our sample. We also show the modelled beam-smearing contribution to the dispersion field as well as the beam-smearing corrected dispersion field obtained using the method described above. We also show the observed and beam-smearing corrected rotation curve $V_\mathrm{CO}$ and the $V/\mathrm{\sigma}_\mathrm{CO}$ ratio.

This method serves as an estimation of the beam-smearing effect and may not fully capture the beam-smearing effect because we utilise only the mean velocity map, instead of applying beam-smearing correction to each and every channel. Also, the underlying gas distribution is assumed to be uniform, which is not necessarily the case. In addition, we cannot take into account any seeing effects with this method. Nevertheless, we can see that for most of the observed dispersion field, the patterns that are caused by the beam-smearing effect can be reproduced in the modelled dispersion field, and hence be subtracted.

\section{Possible effects of $\lowercase{m}=2$ perturbation on $V_\mathrm{CO}$}\label{app_c1p}
\begin{figure}
\includegraphics[trim=0 0 0 150, clip=true,width=0.48\textwidth]{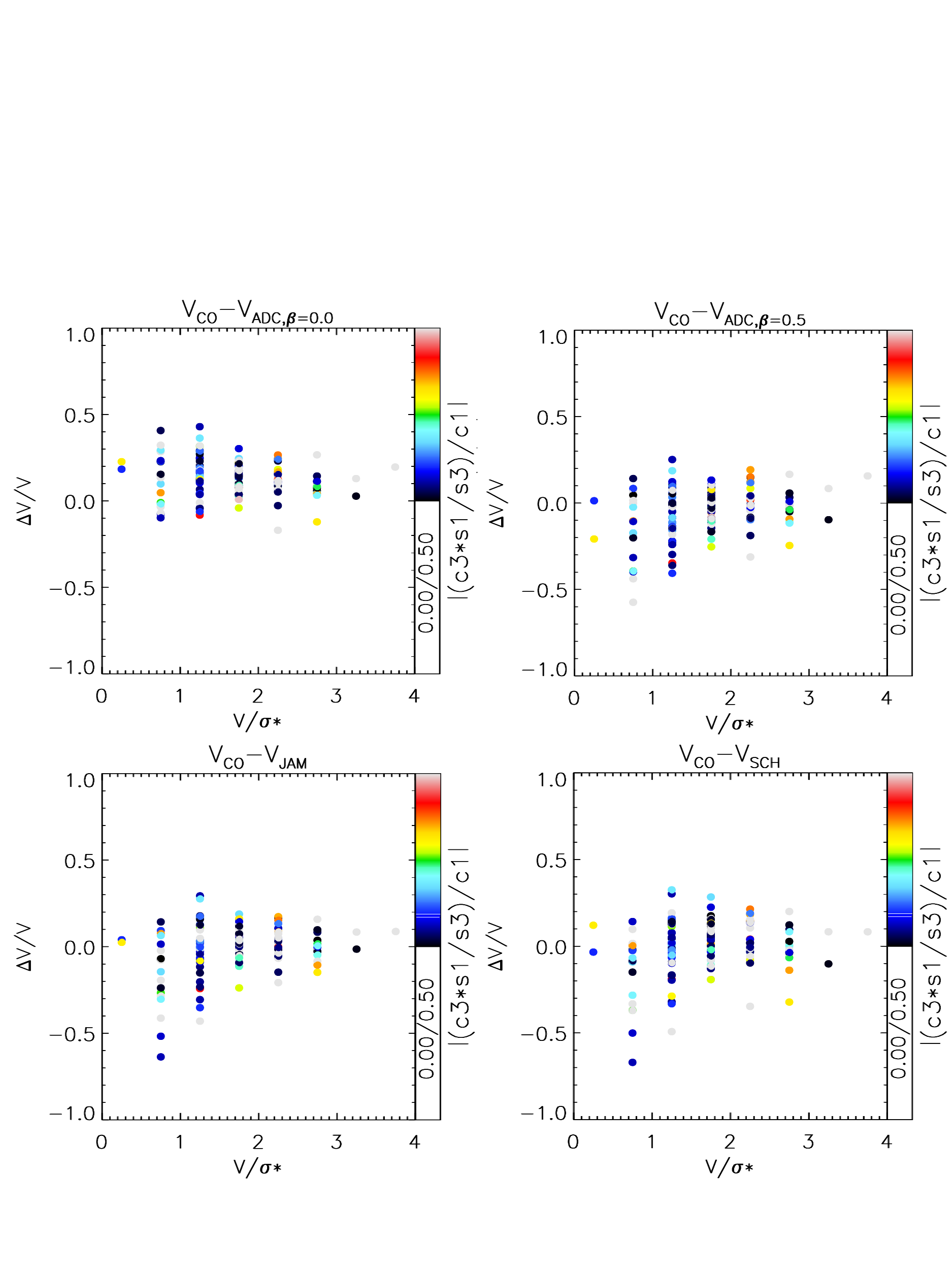}
\caption{$\Delta V/V$ plotted against $V/\sigma_\star$, colour coded with the corresponding $|(c_{3}s_{1}/s_{3})/c_{1}|$ value for each galaxy in the specific $V/\sigma_\star$ bin. No trends in $\Delta V/V$ are seen with respect to $|(c_{3}s_{1}/s_{3})/c_{1}|$.}
\label{fig_c1p}
\end{figure}

We extracted $V_\mathrm{CO}$ with harmonic decomposition:
\begin{equation}
\begin{aligned}
&V_\mathrm{mod} = V_\mathrm{sys} + c_{1}\cos(\phi) + s_{1} \sin(\phi) + c_{2}\cos(2\phi) \\
&+ s_{2}\sin(2\phi)+ c_{3}\cos(3\phi) + s_{3}\sin(2\phi),
\end{aligned}
\end{equation}
From here, we take $c_{1}/\sin(i)$ as $V_\phi=V_\mathrm{CO}$. In fact, although most of the high-order perturbation can be removed using this method, perturbation of $m=2$ mode can still have an effect on on $c_{1}$. As described in \citet{spe07}, the effect on m=2 mode perturbation on $c_{1}$ can be estimated as $c_{1}=V_\phi+c_{3}(s_{1}-V_\mathrm{rad})/s_{3}$, where $V_\mathrm{rad}$ is the first order radial flow. All the galaxies in our sample have average $s_{1}$, $c_{3}$ and $s_{3}$ terms of $\lesssim$10\% of $c_{1}$. While we do not have independent handle on $V_\mathrm{rad}$, $s_{1}$ in general should be dominated by radial flow such that $s_{1}\sim V_\mathrm{rad}$. To put an upper limit on how much $c_{1}/\sin(i)$ deviate from $V_\phi$, we assume that $s_{1}$ is completely dominated by $m=2$ perturbation, i.e. $V_\mathrm{rad}=0$. In Figure \ref{fig_c1p}, we plot for each stellar dynamical model, $\Delta V/V$ versus $V/\sigma_\star$ (as in Figure \ref{fig_global_vod}), colour coded with the corresponding $|(c_{3}s_{1}/s_{3})/c_{1}|$ value for each galaxy in the specific $V/\sigma_\star$ bin. $|c_{3}s_{1}/s_{3}|$ gives an upper limit to how much $c_{1}/\sin(i)$ deviate from the true $V_\phi$. We show here the high $\Delta V/V$ points for each models in the low $V/\sigma_\star$ regime are not caused by possible contribution of higher order perturbation in $V_\mathrm{CO}$ as the corresponding points have low $|(c_{3}s_{1}/s_{3})/c_{1}|$ values. The large scatters in $\Delta V/V$ in the low $V/\sigma_\star$ regime are also not caused by higher order perturbations as there are no trends seen with respect to $|(c_{3}s_{1}/s_{3})/c_{1}|$.

\section{Issues with unphysical parameters with JAM}\label{app_jam}
As discussed in Section \ref{subsect_jam}, 7 galaxies in our sample converge towards the boundary condition of $\beta_\mathrm{z}=-2$ and 7 other galaxies converge towards the boundary condition of $V_\mathrm{vir}=400$\,km\,s$^{-1}$ when a stellar vs. halo mass condition is not applied. We quantify here how such unphysical solutions affect our results. 

\begin{figure*}
\includegraphics[trim=0 20 50 50, clip=true,width=0.9\textwidth]{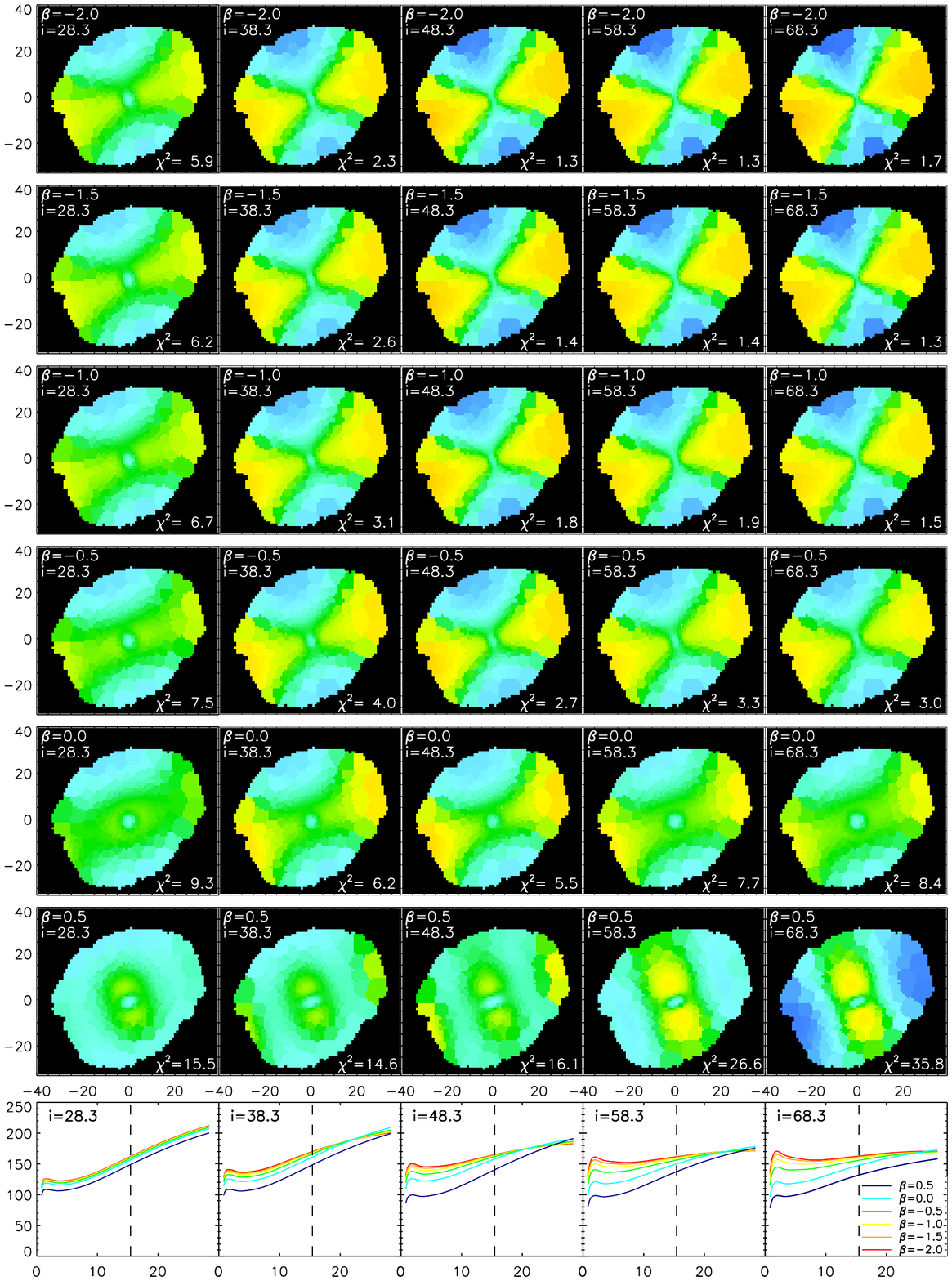}
\caption{Best-fitted kinematic maps with fixed $\beta_\mathrm{z}$ and inclination $i$. Individual panels in the bottom row shows the derived $V_\mathrm{c}$ for different fixed inclinations, as marked on the top left corner of each panel. At the bottom right corner, we show the reduced $\chi^{2}$ of each model. Within each panel, $V_\mathrm{c}$ derived with different fixed $\beta_\mathrm{z}$ are plotted with different colours. Vertical dashed lines mark the effective radius.}
\label{fig_jam_betaincl}
\end{figure*}

We first show that the $\beta_\mathrm{z}<-1.5$ cases (i.e. the 7 galaxies marked with $\dagger$ in Table \ref{tab_jam}) are not merely caused by an incorrect inclination estimate. As an example we show in Figure \ref{fig_jam_betaincl}, the best-fitted $V_\mathrm{rms}$ maps at fixed $\beta_\mathrm{z}$ of [$-2.0,\,-1.5,\,-1.0,\,-0.5,\,0.0,\,0.5$] and vary the inclination with respective to $i=48.3\degree$ (as derived from the ellipticity of the outer isophotes of r-band photometry) with $\Delta i$ of [$-20\degree,\,-10\degree,\,0\degree,\,10\degree,\,20\degree$]. In every point of the grid ($\beta_\mathrm{z}$, $i$) are fixed, but ($\Upsilon_\star$, $V_\mathrm{vir}$) are free parameters. The value of the best-fit ($\beta_\mathrm{z}$, $i$) of individual galaxies are determined by the shape of the $V_\mathrm{rms}$ map. There are degeneracies between ($\beta_\mathrm{z}$, $i$), in the sense that a more negative $\beta_\mathrm{z}$ and a higher $i$ have similar effects on the shape of the $V_\mathrm{rms}$ field. We find that for the 7 galaxies marked with $\dagger$ however, even with a $\Delta i$ of 20$\degree$, the best fitted model still have $\beta_\mathrm{z}\leq-1.5$. This suggest that the low $\beta_\mathrm{z}$ values we find are not just an effect of an incorrectly estimated inclinations, but are intrinsic to the JAM models.

We also show how different $\beta_\mathrm{z}$ and $i$ value affect the derived $V_\mathrm{JAM}$ on the bottom row of Figure \ref{fig_jam_betaincl}. For $\beta_\mathrm{z}<-0.5$, $V_\mathrm{JAM}$ agree to within $\sim$1\% at 1\,$R_\mathrm{e}$ for any inclinations, suggesting that a highly negative $\beta_\mathrm{z}$ has only negligible effect on the derived $V_\mathrm{c}$. The $V_\mathrm{JAM}$ derived also provide good agreement with $V_\mathrm{CO}$. We therefore do not impose further constrain on $\beta_\mathrm{z}$. Restricting $\beta_\mathrm{z}>0$ for example, on the other hand, would change the shape of the derived $V_\mathrm{c}$ to deviate from $V_\mathrm{CO}$ and therefore we do not suggest such practice.

\begin{figure}
\includegraphics[trim=0 0 0 0, clip=true,width=0.6\textwidth]{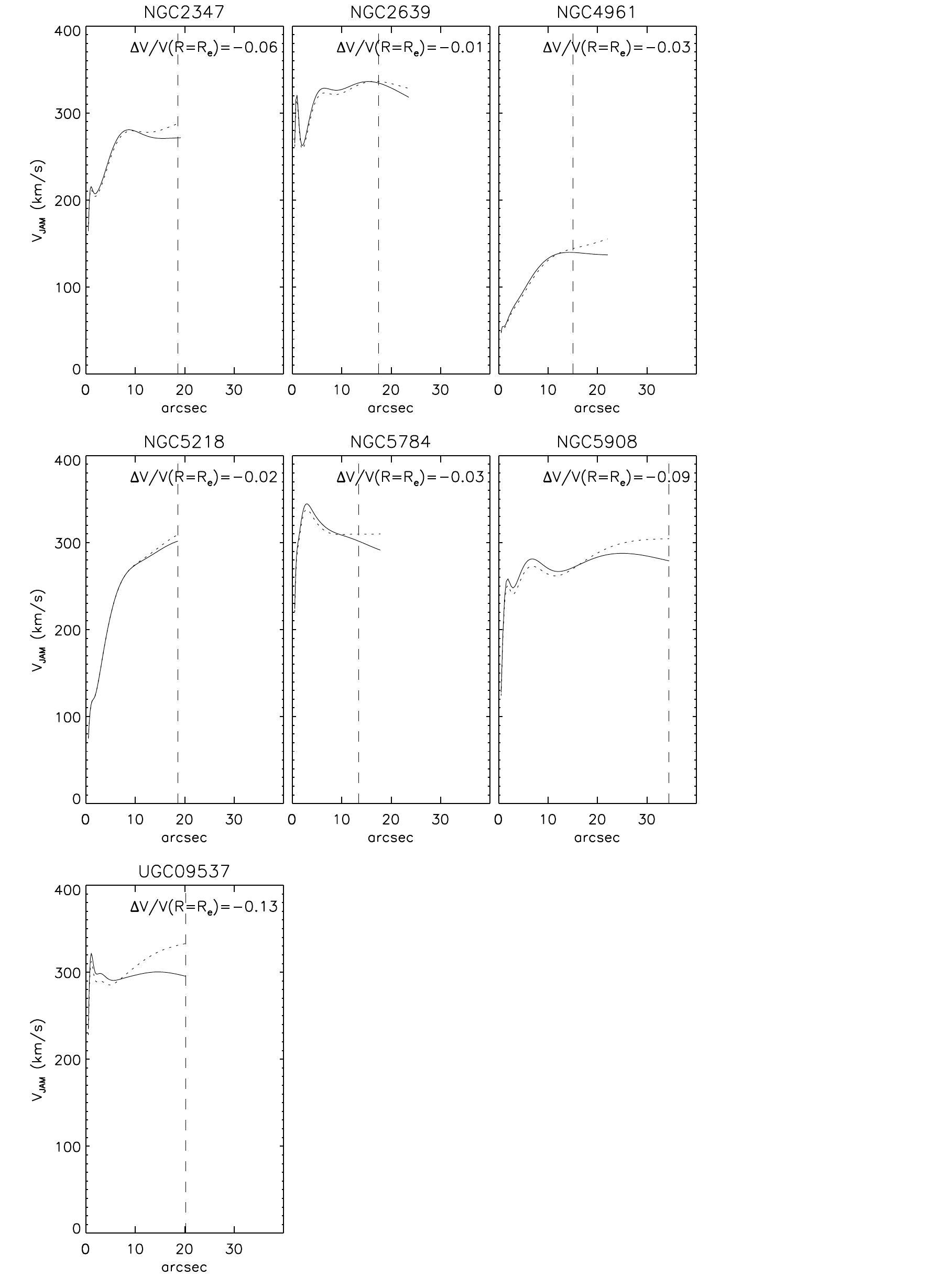}
\caption{$V_\mathrm{JAM}$ for the 7 galaxies marked with $\ddagger$ in Table \ref{tab_jam} between when we impose the stellar-mass-halo-mass relation (solid lines) and when we impose an uniform prior of $0-400$\,km\,s$^{-1}$ to $V_\mathrm{vir}$ (dotted lines). On the top right corner of each panel, we show the relative difference between the two $V_\mathrm{JAM}$ at 1\,$R_\mathrm{e}$.}
\label{fig_jam400re}
\end{figure}

We show in Figure \ref{fig_jam400re} $V_\mathrm{JAM}$ for the 7 galaxies which has $V_\mathrm{vir}$ driven to the upper boundary of 400\,km\,s$^{-1}$ (marked with $\ddagger$ in Table \ref{tab_jam}). The best fit $V_\mathrm{c}$ when we impose a uniform prior of $0<V_\mathrm{vir}<400$\,km\,s$^{-1}$ is shown in dotted lines. The $V_\mathrm{c}$ in models where we impose an additional stellar-mass-halo-mass relation (Eq. \ref{eq_smhm}) are plotted in solid lines. In 4 of the galaxies, NGC2639, NGC4961, NGC5218 and NGC5784, the differences between the two $V_\mathrm{c}$ are only $\leq$3\%. For the other 3 galaxies, NGC2347, NGC5908, and UGC09537, however, $V_\mathrm{JAM}$ shows a steep rise towards large radii. Such steep rises suggest that an unphysically high $V_\mathrm{vir}$ can have an effect on the derived $V_\mathrm{c}$ and therefore it is necessary to impose Eq. \ref{eq_smhm} to galaxies which do not have $V_\mathrm{vir}$ converging within the imposed prior.


\section{Observed and modelled stellar photometry and kinematics}\label{app_stellarmod}
\bsp
\newpage
\begin{figure*}
\centering
\includegraphics[trim=73 40 55 20, page=1,width=0.6\textwidth,angle=90]{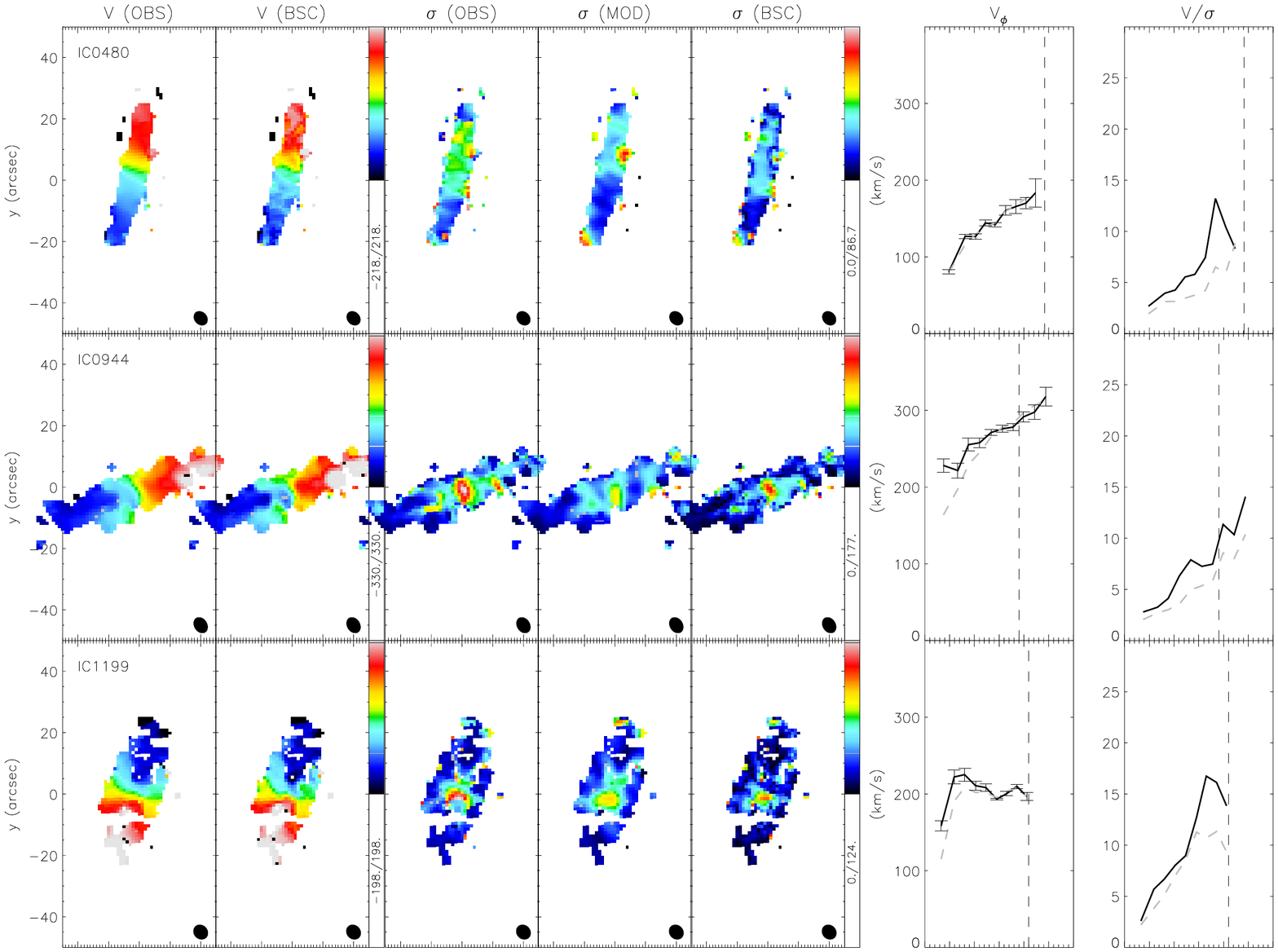}
\includegraphics[trim=100 40 28 20, page=2,width=0.6\textwidth,angle=90]{plotvod_all.pdf}
\end{figure*}
\begin{figure*}
\centering 
\includegraphics[trim=73 40 55 20, page=3,width=0.6\textwidth,angle=90]{plotvod_all.pdf}
\includegraphics[trim=100 40 28 20, page=4,width=0.6\textwidth,angle=90]{plotvod_all.pdf}
\end{figure*}
\begin{figure*}
\centering 
\includegraphics[trim=73 40 55 20, page=5,width=0.6\textwidth,angle=90]{plotvod_all.pdf}
\includegraphics[trim=100 40 28 20, page=6,width=0.6\textwidth,angle=90]{plotvod_all.pdf}
\end{figure*}
\begin{figure*}
\centering 
\includegraphics[trim=73 40 55 20, page=7,width=0.6\textwidth,angle=90]{plotvod_all.pdf}
\includegraphics[trim=100 40 28 20, page=8,width=0.6\textwidth,angle=90]{plotvod_all.pdf}
\end{figure*}

\begin{figure*}
\centering 
\includegraphics[trim=73 40 55 20, page=9,width=0.6\textwidth,angle=90]{plotvod_all.pdf}
\includegraphics[trim=100 40 28 20, page=10,width=0.6\textwidth,angle=90]{plotvod_all.pdf}
\end{figure*}
\begin{figure*}
\centering 
\includegraphics[trim=73 40 55 20, page=11,width=0.6\textwidth,angle=90]{plotvod_all.pdf}
\includegraphics[trim=100 40 28 20, page=12,width=0.6\textwidth,angle=90]{plotvod_all.pdf}
\end{figure*}
\begin{figure*}
\centering 
\includegraphics[trim=73 40 55 20, page=13,width=0.6\textwidth,angle=90]{plotvod_all.pdf}
\includegraphics[trim=100 40 28 20, page=14,width=0.6\textwidth,angle=90]{plotvod_all.pdf}
\end{figure*}
\begin{figure*}
\centering 
\includegraphics[trim=73 40 55 20, page=15,width=0.6\textwidth,angle=90]{plotvod_all.pdf}
\includegraphics[trim=100 40 28 20, page=16,width=0.6\textwidth,angle=90]{plotvod_all.pdf}
\end{figure*}

\begin{figure*}
\centering
\includegraphics[trim=120 40 150 60, page=1,width=0.5\textwidth,angle=90]{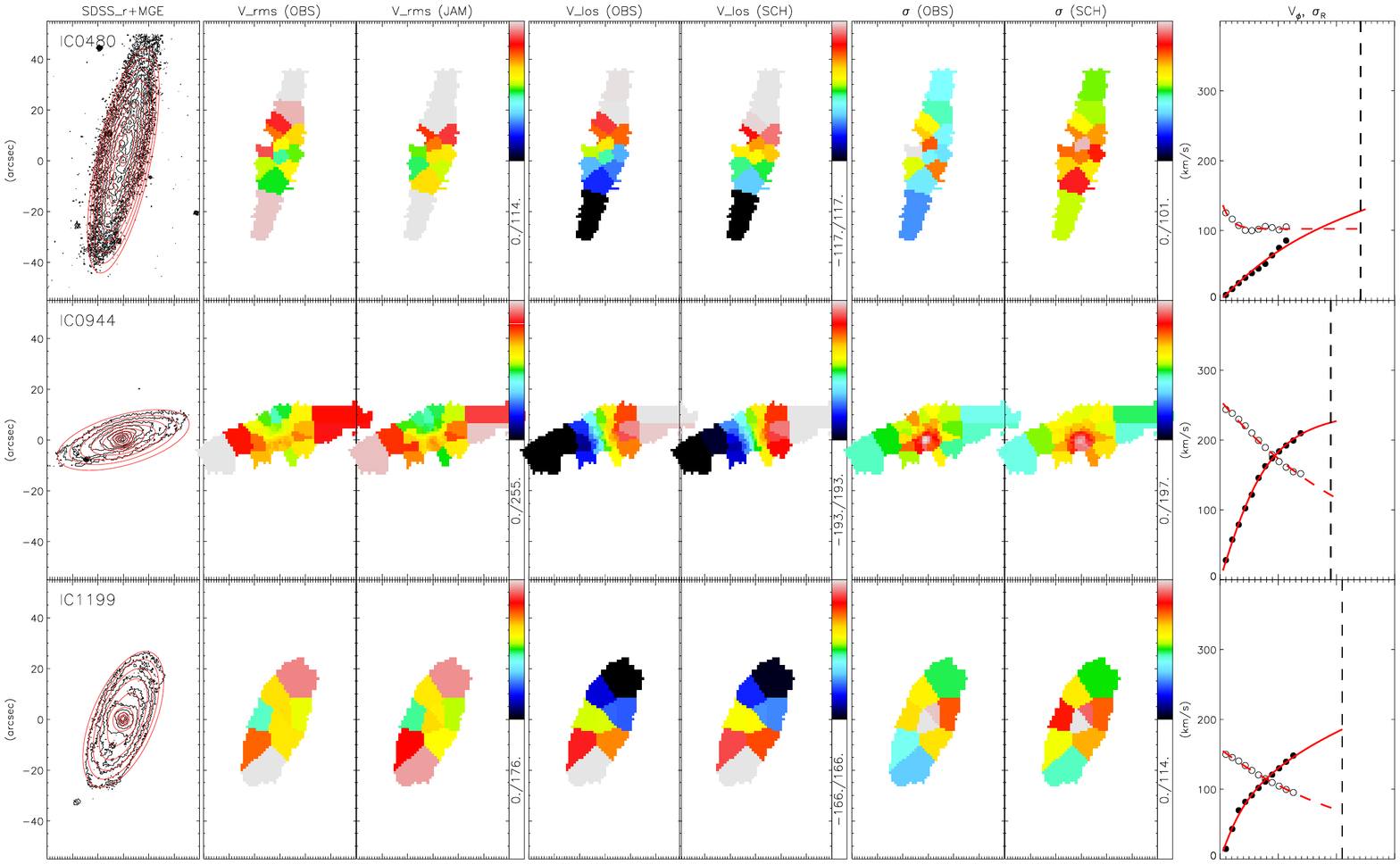}
\includegraphics[trim=207 40 63 60, page=2,width=0.5\textwidth,angle=90]{kinmod_all.pdf}
\end{figure*}
\begin{figure*}
\centering 
\includegraphics[trim=120 40 150 60,  page=3,width=0.5\textwidth,angle=90]{kinmod_all.pdf}
\includegraphics[trim=207 40 63 60, page=4,width=0.5\textwidth,angle=90]{kinmod_all.pdf}
\end{figure*}
\begin{figure*}
\centering 
\includegraphics[trim=120 40 150 60,  page=5,width=0.5\textwidth,angle=90]{kinmod_all.pdf}
\includegraphics[trim=207 40 63 60, page=6,width=0.5\textwidth,angle=90]{kinmod_all.pdf}
\end{figure*}
\begin{figure*}
\centering 
\includegraphics[trim=120 40 150 60,  page=7,width=0.5\textwidth,angle=90]{kinmod_all.pdf}
\includegraphics[trim=207 40 63 60, page=8,width=0.5\textwidth,angle=90]{kinmod_all.pdf}
\end{figure*}
\begin{figure*}
\centering 
\includegraphics[trim=120 40 150 60,  page=9,width=0.5\textwidth,angle=90]{kinmod_all.pdf}
\includegraphics[trim=207 40 63 60, page=10,width=0.5\textwidth,angle=90]{kinmod_all.pdf}
\end{figure*}
\begin{figure*}
\centering 
\includegraphics[trim=120 40 150 60,  page=11,width=0.5\textwidth,angle=90]{kinmod_all.pdf}
\includegraphics[trim=207 40 63 60, page=12,width=0.5\textwidth,angle=90]{kinmod_all.pdf}
\end{figure*}
\begin{figure*}
\centering 
\includegraphics[trim=120 40 150 60,  page=13,width=0.5\textwidth,angle=90]{kinmod_all.pdf}
\includegraphics[trim=207 40 63 60, page=14,width=0.5\textwidth,angle=90]{kinmod_all.pdf}
\end{figure*}
\begin{figure*}
\centering 
\includegraphics[trim=120 40 150 60,  page=15,width=0.5\textwidth,angle=90]{kinmod_all.pdf}
\includegraphics[trim=207 40 63 60, page=16,width=0.5\textwidth,angle=90]{kinmod_all.pdf}
\end{figure*}
\begin{figure*}
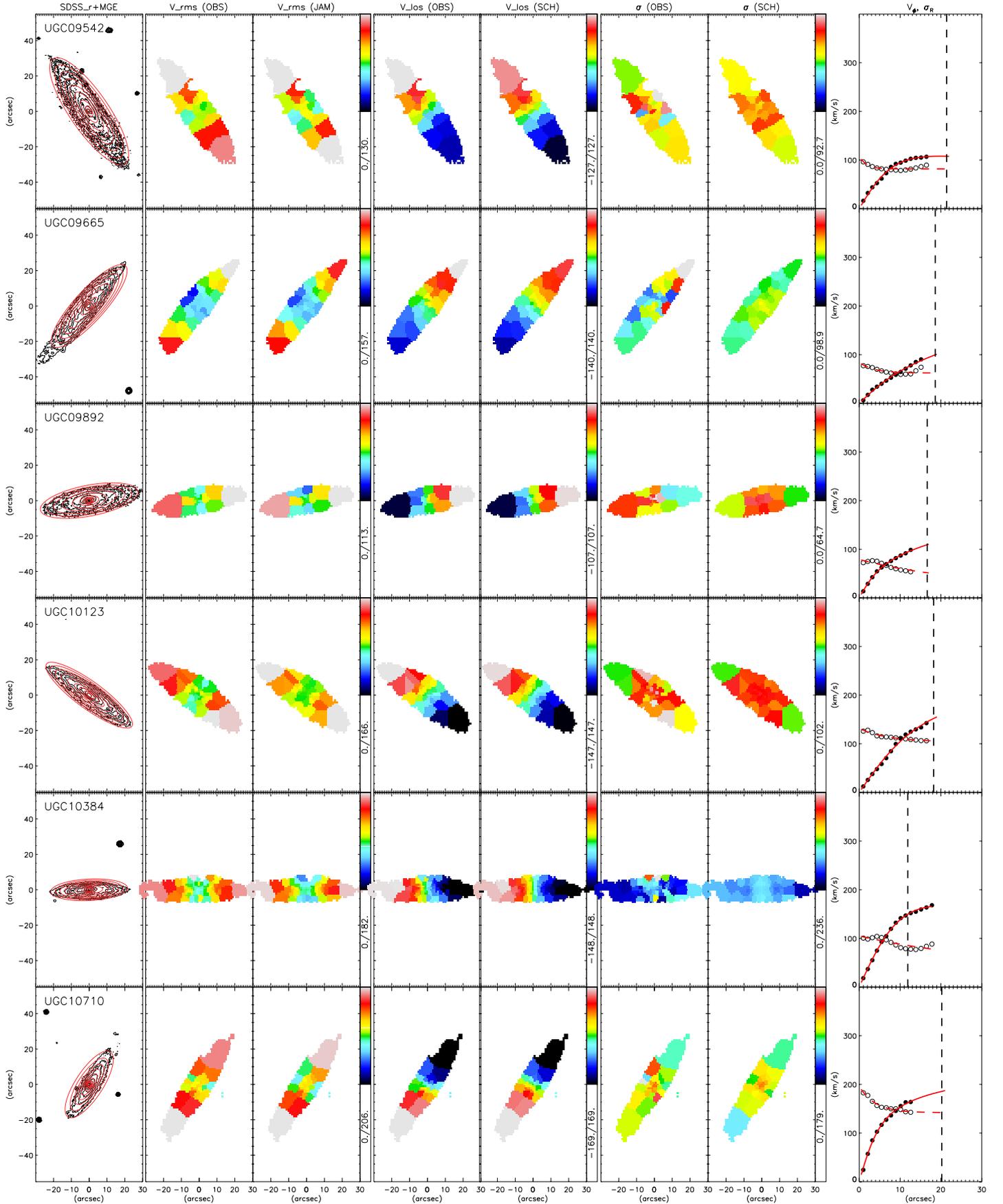

\centering 
\includegraphics[trim=120 40 150 60,  page=17,width=0.5\textwidth,angle=90]{kinmod_all.pdf}
\includegraphics[trim=207 40 63 60, page=18,width=0.5\textwidth,angle=90]{kinmod_all.pdf}
\vspace*{50mm}
\caption{The eight figures for each galaxy from left to right are: (1) observed SDSS r-band image in black and the fitted MGEs over-plotted in red contours, (2) observed $V_\mathrm{rms}$, (3) best-fitted JAM $V_\mathrm{rms}$, (4) observed $V_\mathrm{los}$, (5) best-fitted Schwarzschild modelled $V_\mathrm{los}$, (6) observed $\sigma$, (7) best-fitted Schwarzschild modelled $\sigma$, all colour coded in scales of km\,s$^{-1}$ as denoted by the colour bars; (8) the extracted observed kinematics: $V_\phi$ and $\sigma_R$ plotted in filled and open circles respectively. The fitted form used in ADC as mentioned in the main text are over-plotted in solid red lines for $V_\phi$ and red dashed lines for $\sigma_R$ (for $\beta=0.5$). The vertical dashed line represent 1\,$R_{e}$.}
\label{fig_kinmodall}
\end{figure*}

\label{lastpage}

\end{document}